\documentclass[%
  reprint,
 amsmath,amssymb,longbibliography
 aps,
prd]{revtex4-2}

\usepackage{graphicx}
\usepackage{dcolumn}
\usepackage{bm}
\usepackage[mathlines]{lineno}
\usepackage{lipsum}
\usepackage{siunitx}
\usepackage{glossaries}
\newacronym{AT}{AT}{Arnold tongue}
\newacronym{RF}{RF}{radio-frequency}
\newacronym{PN}{PN}{phase noise}
\newacronym{OMO}{OMO}{optomechanical oscillator}
\newacronym{TE}{TE}{transverse-electric}
\newacronym{ESA}{ESA}{electrical spectrum analyzer}
\newacronym{sinf}{sinf}{Supplementary Note}
\newacronym{KBM}{KBM}{Krylov-Bogoliubov-Mitropolsky}
\usepackage{todonotes}
\usepackage{dirtytalk} 
\usepackage{hyperref}
\hypersetup{
    colorlinks=true,
    linkcolor=red,
    anchorcolor=black,
    citecolor=blue,
    filecolor=cyan,
    menucolor=red,
    runcolor=cyan,
    urlcolor=magenta,
    pdftitle={Optomechanical Synchronization across Multi-Octaves Frequency Spans},
    pdfpagemode=FullScreen,
    }
\usepackage{cleveref}
\Crefname{equation}{Eq.}{Eqs.}
\Crefname{figure}{Fig.}{Figs.}
\Crefname{tabular}{Tab.}{Tabs.}

\begin{document}

\preprint{APS/123-QED}

\title{Optomechanical Synchronization across Multi-Octave Frequency Spans}

\author{Caique C. Rodrigues$^{1,2}$}
\affiliation{\vspace{0.25cm} $^1$Applied Physics Department, Gleb Wataghin Physics Institute, University of Campinas, Campinas, SP, Brazil \\
$^2$Photonics Research Center, University of Campinas, Campinas, SP, Brazil\\
$^3$Department of Electrical Engineering, Columbia University, New York, New York 10027, USA\\
$^4$Department of Applied Physics and Applied Mathematics, Columbia University, New York, New York 10027, USA}%

\author{Cauê M. Kersul$^{1,2}$}
\affiliation{\vspace{0.25cm} $^1$Applied Physics Department, Gleb Wataghin Physics Institute, University of Campinas, Campinas, SP, Brazil \\
$^2$Photonics Research Center, University of Campinas, Campinas, SP, Brazil\\
$^3$Department of Electrical Engineering, Columbia University, New York, New York 10027, USA\\
$^4$Department of Applied Physics and Applied Mathematics, Columbia University, New York, New York 10027, USA}%

\author{André G. Primo$^{1,2}$}
\affiliation{\vspace{0.25cm} $^1$Applied Physics Department, Gleb Wataghin Physics Institute, University of Campinas, Campinas, SP, Brazil \\
$^2$Photonics Research Center, University of Campinas, Campinas, SP, Brazil\\
$^3$Department of Electrical Engineering, Columbia University, New York, New York 10027, USA\\
$^4$Department of Applied Physics and Applied Mathematics, Columbia University, New York, New York 10027, USA}%

\author{Michal Lipson$^{3,4}$}
\affiliation{\vspace{0.25cm} $^1$Applied Physics Department, Gleb Wataghin Physics Institute, University of Campinas, Campinas, SP, Brazil \\
$^2$Photonics Research Center, University of Campinas, Campinas, SP, Brazil\\
$^3$Department of Electrical Engineering, Columbia University, New York, New York 10027, USA\\
$^4$Department of Applied Physics and Applied Mathematics, Columbia University, New York, New York 10027, USA}%

\author{Thiago P. M. Alegre$^{1,2}$}
\affiliation{\vspace{0.25cm} $^1$Applied Physics Department, Gleb Wataghin Physics Institute, University of Campinas, Campinas, SP, Brazil \\
$^2$Photonics Research Center, University of Campinas, Campinas, SP, Brazil\\
$^3$Department of Electrical Engineering, Columbia University, New York, New York 10027, USA\\
$^4$Department of Applied Physics and Applied Mathematics, Columbia University, New York, New York 10027, USA}%

\author{Gustavo S. Wiederhecker$^{1,2}$}
\email{gsw@unicamp.br}
\affiliation{\vspace{0.25cm} $^1$Applied Physics Department, Gleb Wataghin Physics Institute, University of Campinas, Campinas, SP, Brazil \\
$^2$Photonics Research Center, University of Campinas, Campinas, SP, Brazil\\
$^3$Department of Electrical Engineering, Columbia University, New York, New York 10027, USA\\
$^4$Department of Applied Physics and Applied Mathematics, Columbia University, New York, New York 10027, USA}%


\begin{abstract}
      \textbf{Abstract:} Experimental exploration of synchronization in scalable oscillator microsystems has unfolded a deeper understanding of networks, collective phenomena, and signal processing. Cavity optomechanical devices have played an important role in this scenario, with the perspective of bridging optical and radio frequencies through nonlinear classical and quantum synchronization concepts. In its simplest form, synchronization occurs when an oscillator is entrained by a signal with frequency nearby the oscillator's tone, and becomes increasingly challenging as their frequency detuning increases. Here, we experimentally demonstrate entrainment of a silicon-nitride optomechanical oscillator driven up to the fourth harmonic of its \SI{32}{\mega\hertz} fundamental frequency. Exploring this effect, we also experimentally demonstrate a purely optomechanical RF frequency divider, where we performed frequency division up to a 4:1 ratio, i.e., from \SI{128}{\mega\hertz} to \SI{32}{\mega\hertz}.  Further developments could harness these effects towards frequency synthesizers, phase-sensitive amplification and nonlinear sensing.
\end{abstract}

\maketitle

\begin{figure*}[t]
    \includegraphics[width=\linewidth]{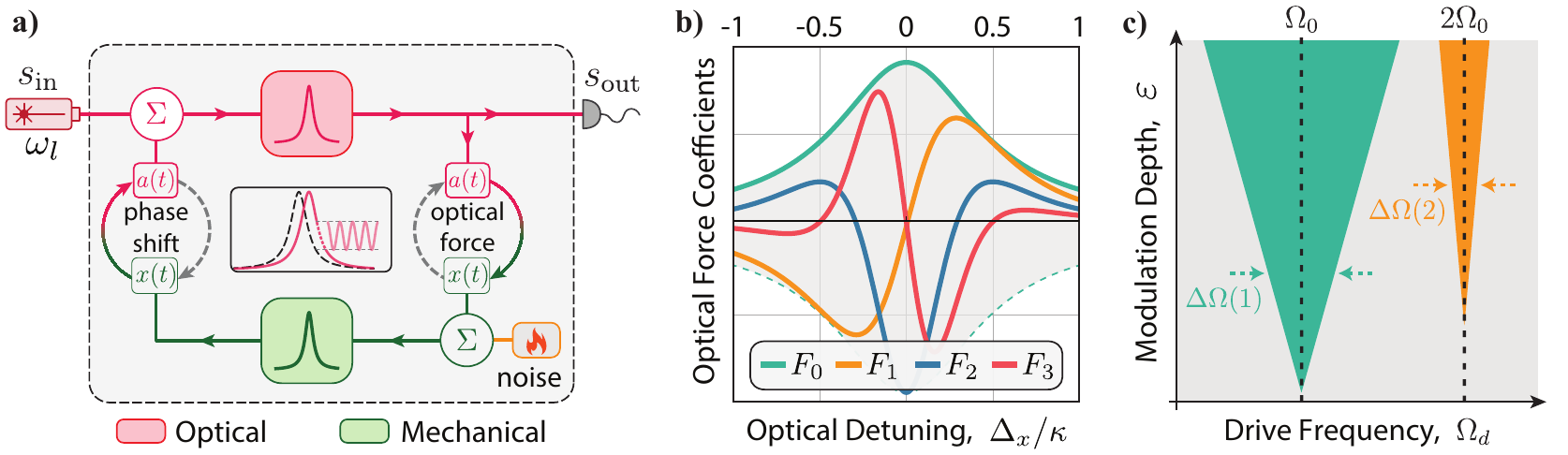}
    \caption{\textbf{High-harmonic response of optomechanical oscillators.} \textbf{a)} Optomechanical oscillator feedback diagram. The mechanical degree of freedom, $x(t)$, is initially in equilibrium with the thermal Brownian noise bath, but when a continuous-wave laser excites the optical field within the optical resonator, $a(t)$, the optical phase is imparted by the mechanical motion and transduced -- via the optical resonance -- to fluctuations on the optical energy. Due to radiation-pressure forces, the mechanical oscillator experiences a feedback (back-action) force that impacts its dynamics; \textbf{b)} Optical force components as function of the optical detuning $\Delta_x = \omega_l - \omega_0 - Gx_0$ shown in \Cref{eq:force_x} (details in the Supplementary Note 5); \textbf{c)} Arnold tongues in the $\varepsilon-\Omega_d$ space illustrating  $1:1$ and $2:1$ entrainment.}
    \label{fig:1}
\end{figure*}

\section{Introduction}

Synchronization lies at the core of time keeping and underpins a vast class of natural phenomena, from life cycles to precision measurements~\cite{Pikovsky2002}. In a nutshell, synchronization occurs when an oscillatory system has its bare frequency entrained by a weak external signal, which may have a slightly different tempo. Since its observation by Huygens in the 17$^\text{th}$ century, the synchronization of widely distinct systems has been shown to share remarkably universal features~\cite{Pikovsky2002, Jenkins2013}, fostering its exploration across many disciplines~\cite{Aspelmeyer2014, Strogatz1994NonlinearEngineering, Jackson1991Perspectives1}. With the recent convergence among optical, mechanical and electrical waves using scalable microfabrication technologies, synchronization has emerged as a powerful tool targeted not only at technological applications, such as phase-lock loops (PLLs) in radio-based communications~\cite{Razavi2004a, Plessas2011, Rategh2003SuperharmonicDividers}, but also at developing the fundamentals of chaotic systems~\cite{Barbosa2019}, injection locking~\cite{Shi2019,Markovic2019a,ArreguiColombanoMairePitantiCapujGriolMartinezSotomayorTorresNavarroUrrios+2021+1319+1327}, electro and optomechanical devices~\cite{Mahboob2012,Huang2018a,Bekker2017InjectionDevice,Huang2018, Sheng2020, Xu2019b, Colombano2019, Bagheri2013PhotonicOscillators}, nonlinear dynamics~\cite{Zhou2019, Huan2019,  Ganesan2019, Parlitz1997, Zou2019, Leijssen2017}, network coupling~\cite{Cabot2017, Matheny2019, Sanavio2020, Raeisi2020}, and quantum synchronization~\cite{Walter2014QuantumOscillator, Lorch2016, Lorch2017, Qiao2020, Roulet2018, Eshaqi-Sani2020}.

Most synchronization realizations occur when the oscillation frequencies involved are barely dissimilar. This is usually the case because most oscillators rely on an underlying frequency-selective resonant response, e.g., mechanical, electrical, or optical resonance, which drastically suppresses off-resonant excitations. Despite the weak response to such non-resonant signals, oscillators with a strong nonlinearity may also synchronize when the ratio between external driving frequency ($\Omega_d$) and the oscillation frequency ($\Omega_0$) is close to a rational number $\rho = p/q$ called winding number~\cite{Kennedy1989TheFractal}, i.e., the ratio $\Omega_d/\Omega_0=p/q$ with $p,q$ being coprime integers. Indeed, higher order $p:q$ synchronization features have been experimentally observed in a variety of nonlinear systems, from Van der Pol's neon-bulb oscillator~\cite{VANDERPOL1927FrequencyDemultiplication} to modern spin-torque oscillators~\cite{Urazhdin2010, Tortarolo2018, Keatley2016a},  micro-electro-mechanical systems (MEMS)~\cite{Seitner2017, Pu2018, Taheri-Tehrani2019, Houri2019, Du2019, PhysRevLett.52.2277}, delay-coupled lasers~\cite{MartinezAvila2009TimeFeedback,Barbosa2019}, nuclear magnetic resonance laser~\cite{Simonet1994LockingFeedback}, and on-chip optical parametric oscillators~\cite{Jang2019ObservationCombs}. These higher-order synchronization demonstrations are of major importance in radio-frequency (RF) division applications, which often demand low-power consumption and wide-band operation~\cite{Rocheleau2014AReduction,Amann2009MechanismDividers,Kennedy2011ExperimentalInjection}. 

Within optomechanical devices, although seminal work has revealed that high-order synchronization is possible, its full strength is yet to be developed, potentially impacting the bridge between optical and RF signals~\cite{Hill2012} or enabling role in quantum~\cite{Chan2011LaserState,Lorch2017,Kato2019} and classical devices~\cite{Bagheri2013PhotonicOscillators,PhysRevLett.107.043603}. For instance, the first optomechanical injection-locking demonstration by Hossein-Zadeh et al.~\cite{Hossein-Zadeh2008a} showed evidence of synchronization at $\Omega_d=2\Omega_0$, while ~\cite{Wang2016DevilsCavity,PhysRevE.91.032910} demonstrated synchronization at subharmonics and the second harmonic in an on-fiber optomechanical cavity oscillator based on thermal effects. Theoretical work has suggested weak signatures of higher-order synchronization in optomechanical cavities~\cite{Amitai2017}.

Here, we experimentally demonstrate the entrainment of a silicon-nitride optomechanical oscillator (OMO) by an external signal up to two octaves away from its oscillation frequency. Furthermore, the OMO operates in the intriguing regime where higher order synchronization ($p>q$) is actually stronger than the trivial $1:1$ case, as determined by the degree of nonlinearity set by the laser frequency and intensity. Finally, we explore this regime to experimentally demonstrate a purely optomechanical radio-frequency divider with a phase noise performance better than the 1:1 locking regime. Our results open a route for exploring and engineering nonlinear synchronization in optomechanical oscillators~\cite{Qiao2018}, phase-sensitive amplification~\cite{Rugar1991MechanicalSqueezing,Zega2015PredictingOscillators}, nonlinear sensing~\cite{Brawley2016NonlinearMotion}, and collective dynamics of emerging oscillator arrays~\cite{Pelka2020,Zhang2015SynchronizationLight,Raeisi2020}.

\section{Results and Discussion}

The general structure of optomechanical oscillators dynamic can be represented by the feedback diagram shown in \Cref{fig:1}(a). The optical force driving the mechanical mode depends nonlinearly on the displacement, $x(t)$. Thus, the Lorentzian-shape of the optical resonance provides a unique route to tailor the degree of nonlinearity of the optical force, defining how different harmonics of the mechanical oscillation are excited during the optical-to-mechanical transduction. 

To establish synchronization, we apply a weak intensity modulation to the optical driving power, $P_{\text{in}}(t)=P_0\left[1+\varepsilon\sin\left(\Omega_d t\right)\right]$, where $P_0$ is the continuous-wave average power and $\varepsilon$ $(\ll 1)$ is the modulation depth. In the unresolved sideband regime, where $\Omega_0$ is smaller than the optical linewidth $\kappa$, the essence of the feedback loop of \Cref{fig:1}(a) is captured by introducing a delayed mechanical response $x(t) \rightarrow \widetilde{x}(t-\tau)$, where $\widetilde{x}$ is a normalized dimensionless displacement (details in the Supplementary Note 5). The optical force can then be efficiently written as a power series in $\widetilde{x}(t-\tau)$,

\vspace{-0.2cm}
\begin{equation}\label{eq:force_x}
    F_\text{opt}(t) = f_\text{opt}\left[1+\varepsilon \sin\left(\Omega_d t\right)\right]\sum_{n=0}^{\infty} F_n \widetilde{x}^n\left(t-\tau\right),
\end{equation}

\noindent whose strength depends not only on the overall optical force strength, $f_{\text{opt}}$, but also on the dimensionless coefficients $F_n$, which dictates the intensity of the nonlinearity and their detuning dependence, as shown in \Cref{fig:1}(b). Important optomechanical properties, such as optical cooling/amplification or spring effect~\cite{Chan2011LaserState,Marquardt2006}, are described by considering up to the first-order term $F_1$ in \Cref{eq:force_x}. The modulation depth dependent terms $(\propto \varepsilon)$ enable the injection-locking and synchronization of the OMO to an external drive. While $F_{0}$ and $F_{1}$ hardly provide new insights into synchronization properties, the quadratic and cubic terms ($F_2$ and $F_3$) highlight a key aspect explored in this work: nonlinear synchronization properties can be adjusted with an easily accessible parameter, the optical detuning, which significantly changes their relative strengths, as shown in \Cref{fig:1}(b).

The impact of these nonlinearities in the synchronization dynamics can be cast into the well-known Adler's model, which describes the slowly varying phase dynamics of an oscillator perturbed by a weak external drive~\cite{Adler1973,Amitai2017}. Indeed, we show in ``Methods'' that the Taylor-series description of \Cref{eq:force_x} leads to an effective Adler model when the optical modulation frequency is tuned towards a chosen harmonic of the mechanical frequency. Synchronization in this model arises when the perturbation strength overcomes the frequency mismatch between the drive and oscillator's harmonics. As the external drive frequency $\Omega_d$ is swept around the oscillator harmonics, the synchronization condition may still be satisfied and defines a region in a $\varepsilon-\Omega_d$ space known as Arnold tongues (AT)~\cite{Pikovsky2002}, illustrated in \Cref{fig:1}(c). Such response to higher harmonics could be readily explored for radio-frequency division, as we experimentally demonstrate for divisions ratio $2:1$, $3:1$ and $4:1$, the same orders of the measured Arnold tongues maps.

\begin{figure*}
    \centering
    \includegraphics[width=\linewidth]{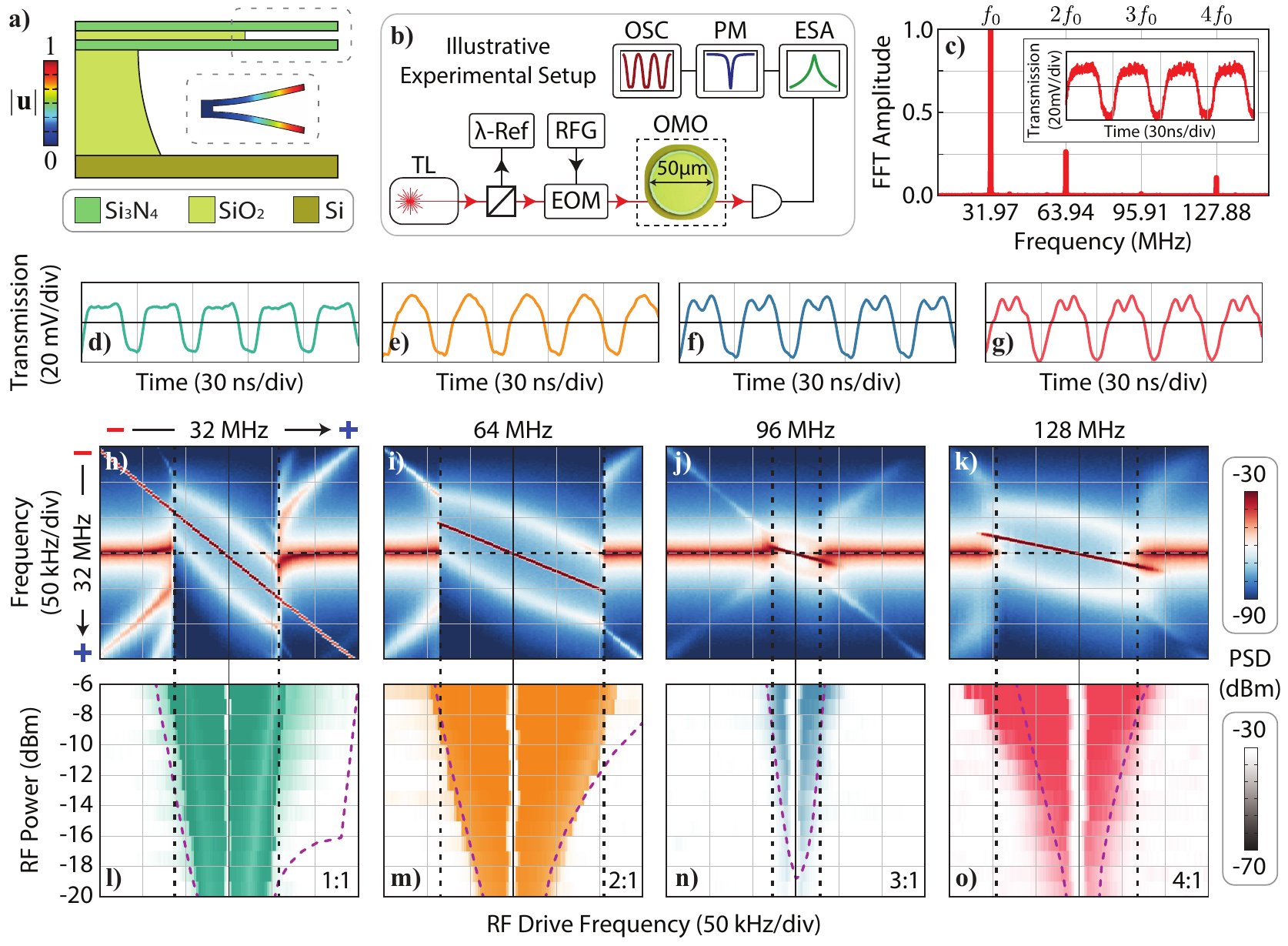}
    \caption{\textbf{Experimental demonstration of multi-octave synchronization.} \textbf{a)} Illustration of the silicon nitride dual-disk optomechanical cavity used in the experiment. The inset shows the simulated flapping mechanical mode displacement profile $|\mathbf{u}|$; \textbf{b)} Schematic of the experimental setup used; TL is the tunable laser source; $\lambda$-Ref: acetylene gas cell and a Mach-Zehnder interferometer as a reference in frequency; EOM: electro-optic modulator; RFG: radio-frequency generator; ESA: electrical spectrum analyzer; PM: power meter; OSC: oscilloscope; \textbf{c)} Magnitude of the fast-Fourier transform of the OMO output signal (inset); \textbf{d)}-\textbf{g)} Time-trace of the OMO output entrained at $p=1$ (\textbf{d}) until $p=4$ (at \textbf{g)}. A RF injection power of -10 dBm ($\varepsilon \approx 4$\%) was used; \textbf{h)}-\textbf{k)} RF spectrograms measured as the RF drive frequency sweeps from lower to higher frequencies around each OMO harmonic, $p=1$ (\textbf{h}) until $p=4$ (at \textbf{k)}, for an injection RF power of -10 dBm. The vertical RF frequency axis is always centered at the mechanical oscillation frequency $\Omega_0/2\pi = 32$ MHz and increases from top to bottom, as the symbols minus and plus from \textbf{h)} suggests. The same is true for the horizontal axis, which the RF drive frequency increases from the left to the right; \textbf{l-o)} Measured Arnold tongues corresponding to each harmonic, obtained by stacking horizontal linecuts along the dashed black line shown in \textbf{h-k)}. The purple curves are the simulated ATs and the color scale of each plot matches the grayscale range shown in the right.}
    \label{fig:2}
\end{figure*}

To experimentally assess high-order synchronization and measure the ATs, it is important to harness the nonlinear response of an OMO. We achieve this control by employing a dual-disk optomechanical cavity based on silicon-nitride~\cite{Zhang2012SynchronizationLight,Shah2015}, as shown schematically in \Cref{fig:2}(a). This cavity supports a relatively low frequency ($\Omega_m/2\pi=$ \SI{31.86}{\mega\hertz}) and high quality factor mechanical mode ($Q_m=1250$)~\cite{Zhang2014EliminatingInterference}, which is coupled to a transverse-electric optical mode ($Q_{\text{opt}}=1.6\times10^5$ at a wavelength $\lambda\approx$ \SI{1556}{\nano\meter}) with an optomechanical coupling rate $g_0/2\pi =$ \SI{16.2}{\kilo\hertz}. The experimental setup, shown in \Cref{fig:2}(b), essentially consists of an intensity-modulated external cavity tunable laser that is coupled to the optomechanical cavity using a tapered fiber~\cite{Zhang2012SynchronizationLight}. The output light is analyzed with an oscilloscope and an  electrical spectrum analyzer (ESA) that reveals the dynamics of the oscillator while monitoring the optical transmission.

To transition this optomechanical cavity into an OMO we raise the pump power to $P_{0} =$ \SI{480}{\micro\watt} and fine-tune its wavelength such that the detuning between the laser frequency and the cavity resonance corresponds to $\Delta_x =  0.35\kappa$ $(\Delta_x/2\pi\approx$ \SI{408}{MHz}), which is inferred by monitoring the optical transmission. A typical OMO free-running output signal and the corresponding  Fourier transform are shown in \Cref{fig:2}(c), revealing the mildly nonlinear characteristic with a few noticeable harmonics. Interestingly, at this detuning, both the $F_0$ and $F_1$ terms in \Cref{eq:force_x} are of similar strength (see \Cref{fig:1}(b)), suggesting that the nonlinear response to an injection signal should be readily observed. To observe injection-locking, the laser intensity modulation is activated, and the modulation frequency is swept around the OMO fundamental frequency or its harmonics ($p=1-4$ and $q=1$). The time-traces in \Cref{fig:2}(d-g) are captured with the injection signal frequency being precisely matched to each harmonic using a RF power of -10 dBm. As the RF driving frequency is detuned from each harmonic, the OMO response is monitored through the RF spectrum centered around the fundamental frequency $\Omega_0/2\pi$, as shown in the density plots of \Cref{fig:2}(h-k). At the left-hand side of these plots, the RF tone is far away from the OMO harmonics and do not synchronize, thus, both oscillator and drive frequencies appear as distinct peaks, accompanied by nonlinear mixing products typical of driven oscillators~\cite{Seitner2017}. When the RF tone approaches a harmonic, a clear transition occurs and a single RF peak emerges, which is one major signature of synchronization. The first striking feature is the observation of strong synchronization for all the driving harmonics, a phenomenon that has not been reported in optomechanical systems. Second, and most important, the width of the synchronization region for $p=2$ and $p=4$ are larger than the fundamental harmonic ($p=1$). It is also remarkable that the $p=3$ synchronization window is relatively small, counterposing the hierarchy among harmonics.

To map the synchronization window into Arnold tongues and understand the role played by the optical modulation depth, we performed the measurements shown in \Cref{fig:2}(h-k) for a range of RF powers, and built the ATs shown in \Cref{fig:2}(l-o). The colored regions indicate a synchronized state and were obtained by stacking RF spectral slices along the OMO frequency, given by the horizontal dashed-lines in \Cref{fig:2}(h-k). It is worth pointing out that the highest RF power (-6 dBm) corresponds to a modulation depth $\varepsilon \approx 6\%$, ensuring a weak perturbation regime. Although the existence of higher order tongues could be anticipated by qualitative analysis of the nonlinear terms in \Cref{eq:force_x}, further theoretical analysis is necessary to precisely picture their nature.

\begin{figure}
    \centering
    \includegraphics[width=\linewidth]{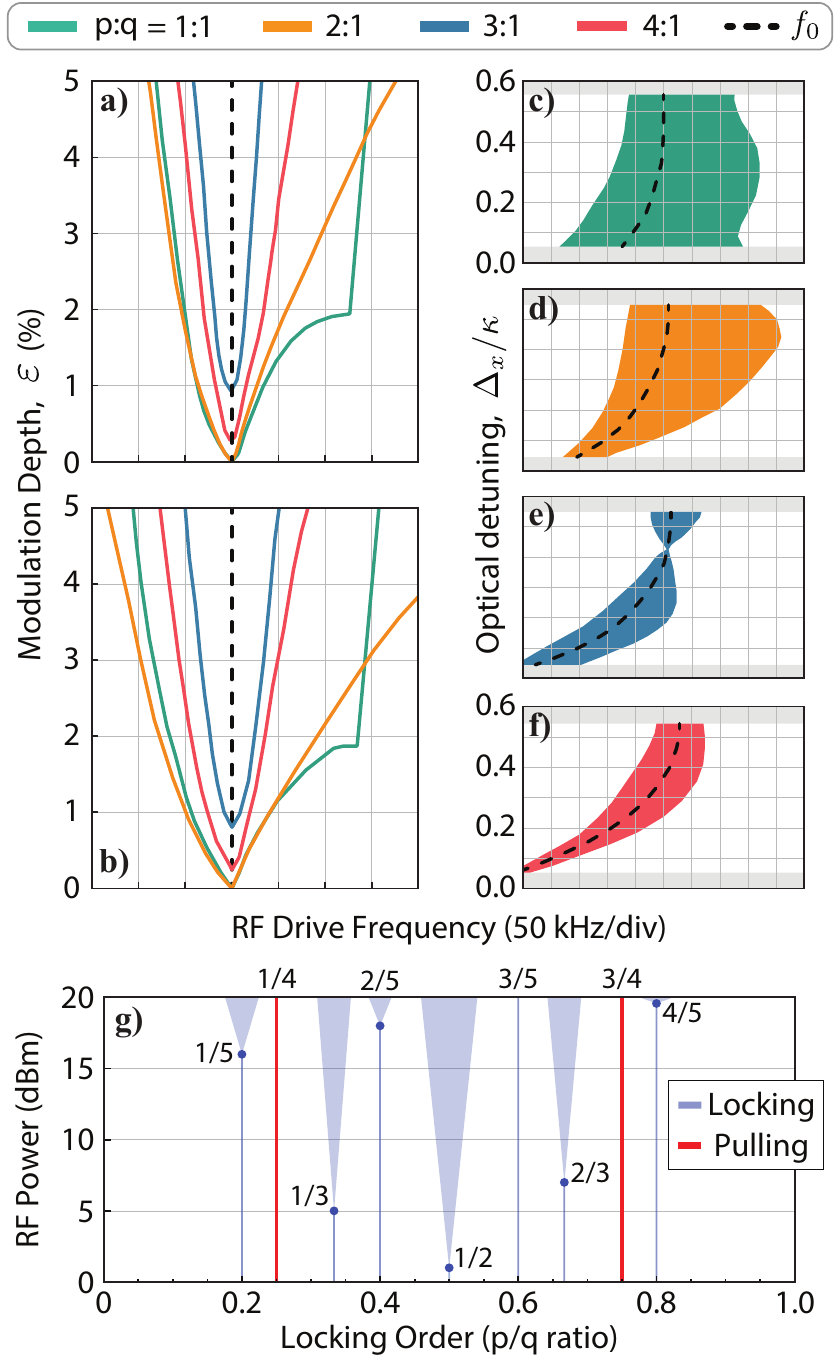}
    \caption{\textbf{Numerical analysis and experimental observation of fractional synchronization.} \textbf{a)} Arnold tongues boundaries simulated using the complete coupled optomechanical equations. The horizontal scale is the same used in experimental data of \Cref{fig:2}\textbf{(l-o)}, revealing a  good agreement; \textbf{b)} Same simulation done at \textbf{a)} but now considering only one parametric term in each simulation, i.e., the green ($1:1$) boundary was simulated considering $\varepsilon F_1 = \varepsilon F_2 = \varepsilon F_3 = 0$ but $\varepsilon F_0 \neq 0$ (details in the Supplementary Note 5). The orange ($2:1$) boundary has only the term $\varepsilon F_1 \neq 0$, the blue ($3:1$) has $\varepsilon F_2 \neq 0$ and the red ($4:1$) has $\varepsilon F_3 \neq 0$; \textbf{c)}-\textbf{f)} Impact of optical detuning $\Delta_x$ in the ATs, showing their tunability and the possibility of a vanishing $p=3$ tongue at $\Delta_x \approx 0.43\kappa$ for the parameters used. These maps were simulated using $\varepsilon = 5\%$ and the black-dashed line is the mechanical oscillation frequency $f_0$, which increases with $\Delta_x$ because of the optical spring effect; \textbf{g)} Measured fractional synchronization threshold, indicated as blue dots, to observe a finite-width AT. The red-lines indicate the locking orders that did not synchronize and only frequency pulling was observed. The Arnold tongues shown are illustrations (see Supplementary Note 2 for actual data).}
    \label{fig:3}
\end{figure}

To study the observed AT behavior, we perform numerical simulations of the exact coupled equations describing both the mechanical and optical dynamics, and the resulting simulated Arnold tongues boundaries are shown in \Cref{fig:3}(a). Despite the specific parameters that influence the precise behavior of the optomechanical limit cycles~\cite{Amitai2017}, such as optical detuning, optomechanical coupling, and optical/mechanical linewidths, a good agreement is observed between the measured and simulated tongues. Such agreement suggests that the observed features are indeed dominated by the optomechanical interaction itself, in contrast to silicon optomechanical devices where thermal and charge carrier effects strongly influence the self-sustaining oscillator dynamics~\cite{Luan2014,Colombano2019}. Although the numerical model is useful for confirming the optomechanical nature of the observed effects, it hardly provides any analytical insight on the origins of the observed synchronization effects. 

We obtain further insight by approximating the optical force as a delayed power series, as suggested in \Cref{eq:force_x}. This analysis allows exploring the synchronization role of each nonlinear component $F_n$ in \Cref{eq:force_x} and elucidates the underlying structure  of high-harmonic synchronization. The nonlinear components that \textit{are not} proportional to the driving signal define a ``forced Van der Pol-Duffing oscillator'' responsible for the oscillator limit cycle observed in \Cref{fig:2}(c). 

The synchronization dynamics is related to the terms proportional to the RF driving signal ($\propto\varepsilon$). However, in addition to the usual non-parametric excitation ($\propto \varepsilon F_0$), the injection signal also contributes to time-dependent coefficients in the mechanical oscillator dynamical equation. Physically, these time-varying coefficients indicate that the external signal modulates the oscillator's frequency and damping properties, leading to linear ($\propto\varepsilon F_1$) and nonlinear ($\propto\varepsilon F_{2,3}$) parametric resonance effects, a situation resembling the dynamics of a nonlinear Mathieu equation~\cite{Kovacic2018MathieusFeatures,Shah2015}.

By neglecting all but one time-dependent term in the numerical simulations, we could identify how each harmonic ($p=1-4$) is related to the force expansion coefficients shown in \Cref{fig:1}(b). The resulting map is shown in \Cref{fig:3}(b), where each boundary was simulated considering only one parametric term, while all the others were set to zero. The resemblance with the full model simulation at \Cref{fig:3}(a) is remarkable. This analysis reveals that each $\varepsilon F_{p-1}$ term in the force expansion is the leading contribution to the $p:1$ AT, for all measured harmonics. For instance, as the $p=3$ entrainment occurs due to the $\varepsilon F_2$ parametric term, the thinner tongue observed in \Cref{fig:2}(n) is explained by the negligible value for $F_2$ at this detuning. Interestingly, although quadratic force terms like $F_2 x^2$ are often ignored in nonlinear mechanical oscillators (as they arise from an asymmetric elastic potential energy), here, they emerge naturally from the Lorentzian shape of the optical mode and can be tuned with the optical detuning. Another interesting feature, present both in the analytical and numerical model, is the presence of a cusp in the 1:1 tongue at -16~dBm RF power. Although we verified that such feature occurs due to an amplitude bifurcation using the analytical model (see Supplementary Note 8), the cusp was not observed in the experimental trace. 

The insights brought by our semi-analytical model suggest that tunable Arnold tongues should be feasible. In \Cref{fig:3}(c-f) we show a full numerical simulation of the ATs as a function the optical detuning, confirming this possibility. In particular, a complete suppression of $p=3$ tongue is attainable (\Cref{fig:3}(e)). Such rich response to higher harmonic excitation led us to verify whether our OMO could also respond to fractional frequency excitation, i.e., where $p/q$ is not an integer number. These experimental results are summarized in \Cref{fig:3}(g) but the full map can be found in the Supplementary Note 2 for various subharmonics of the mechanical frequency, revealing terms of the famous Farey sequence known in number theory \cite{Kennedy1989TheFractal}. Note, however, that the injection signal power required to observe fractional tongues were substantially larger, with some fractions (e.g., 4/5) requiring a full modulation, which is beyond the reach of our semi-analytical approximations ($\varepsilon\approx100\%$).

\begin{figure*}
    \centering
    \includegraphics[width=\linewidth]{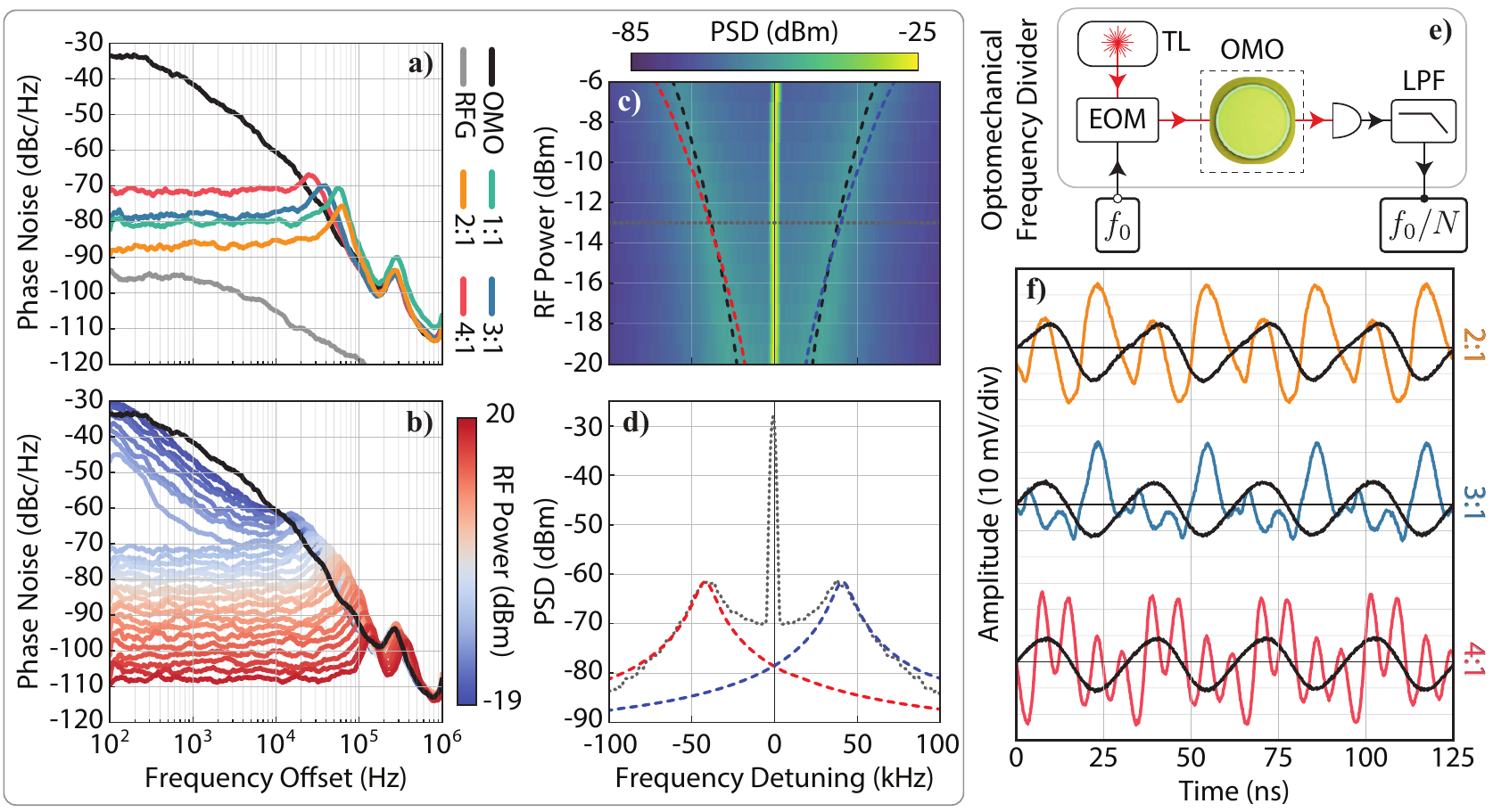}
    \caption{\textbf{Phase-noise reduction and optomechanical frequency division.} \textbf{a)} Measured one-sided phase noise spectral density for the free-running (black), injection-locked (colored) and RF injection signal (gray). The RF power used for the injections was -7 dBm ($\varepsilon\approx 5.5\%$); \textbf{b)} Phase noise spectral density evolution as a function of the RF power for the 4:1 injection; \textbf{c)} Comparison between experimental data sidebands and the semi-analytical model prediction (see Supplementary Note 7). The colormap is the experimental power spectral density (PSD) around the OMO fundamental frequency. The RF drive frequency was set to the OMO frequency for all the RF powers shown. The black dashed lines are the experimental fit and the red/blue curves are the semi-analytical prediction, showing excellent agreement; \textbf{d)} Experimental PSD, in gray, for a RF power of -13 dBm ($\varepsilon \approx 2.75\%$), which is the same horizontal dashed gray linecut at \textbf{c)}, showing the agreement between semi-analytical model and experimental data for these sidebands, for both frequency and linewidth; \textbf{e)} Schematic of the optomechanical frequency divider; LPF means ``low-pass filter"; \textbf{f)} Experimental Optomechanical frequency division. The orange, blue and red curves are the injection-locked signal output from the OMO for the cases 2:1, 3:1 and 4:1, respectively. The overlapping black curves are the divided signal obtained using a low-pass filter with 48 MHz cutoff frequency.}
    \label{fig:4}
\end{figure*}

\textbf{Optomechanical Frequency Division}. An important aspect often praised when investigating synchronization and injection-locking phenomena is the reduction of phase noise (PN) in free-running oscillators. While optomechanical oscillator's phase noise has been previously explored~\cite{Hossein-Zadeh2008a,Fong2014a,Luan2014,Zhang2015SynchronizationLight,Bekker2017InjectionDevice}, its characteristics under high harmonic injection are not known. In \Cref{fig:4}(a) we show the measured PN at the fundamental oscillator's frequency for the free-running OMO and injection-locked at the harmonics $p={1-4}$ (see ``Methods'' for details). The PN curves were taken using a constant RF power of -7 dBm ($\varepsilon\approx 5.5\%$) for all harmonics. The general behavior of the free-running OMO PN has been discussed previously~\cite{Fong2014a} and it is influenced by various noise sources, such as flicker, thermomechanical, and amplitude-to-phase conversion~\cite{Mathai2019}. When injection locked at $p=1$ (green curve), the PN performance improves significantly, and the PN of the higher harmonics are surprisingly low, despite that the same modulation depth was employed. Indeed, the $p=2$ injection offers an improvement over the trivial $p=1$ case, $p=3$ is slightly deteriorated, and $p=4$ PN suffers a significant penalty of 10 dBc/Hz at small offset frequencies, however, it still preserves the low-frequency PN plateau, characteristic of injection-locked oscillators. 
To investigate the RF power dependence of each harmonic, PN curves were measured over a range of RF powers, shown in \Cref{fig:4}(b)  for the 4:1 case. The transition to a low-frequency PN  plateau (around -7 dBm) observed in \Cref{fig:4}(b) also occurs for other harmonics, albeit at lower injection powers, showing that very low PN levels can be achieved at the expense of higher RF power levels (see Supplementary Note 3 for other harmonics). In particular, while the 1:1 PN of \Cref{fig:4}(a) reaches -80 dBc/Hz at -7 dBm, the 4:1 PN of \Cref{fig:4}(b) requires -1 dBm to reach -80 dBc/Hz, still corresponding to a moderate modulation depth of 11\%. A qualitative understanding of the observed PN behavior can be cast upon previous investigations in the context of superharmonic injection-locking~\cite{Zhang1992AOscillators,Verma2003ADividers,Kalia2011,Plessas2011}. When the injection-signal PN is negligible, the phase-noise of a superharmonic injected oscillator is written as

\begin{equation}
    \mathcal{L}_{\text{out}}(\Omega) = \frac{ \mathcal{L}_\text{free}(\Omega)}{1+(\Delta\Omega_n/\Omega)^2\cos^2\theta},
    \label{eq:PN}
\end{equation}

\noindent where $\mathcal{L}_\text{free}(\Omega)$ is the free-running OMO PN spectra, i.e., the black curve of \Cref{fig:4}(a); $\Delta\Omega_n$ is the locking range (AT width) for each harmonic; $\theta$ is the phase offset between the injection signal and the OMO. Apart from the phase offset $\theta$, the AT width determines the locking range and is often associated with good phase noise performance. 

Indeed, the wider lock range $\Delta\Omega_2$ observed for the 2:1 injection is associated with a better PN. For the 3:1 and 4:1 PNs cases, however, the trend is not as clear. While the phase-noise is reduced as the lock-range increases (due to higher injection power), the 4:1 PN shown in \Cref{fig:4}(a) is not lower than the 3:1 injection, despite the wider 4:1 tongue. Although it is not clear all the factors contributing to this discrepancy, we verified in numerical simulations that the phase-offset $\theta$ varies among harmonics and could partially contribute to the observed mismatch. One unique factor contributing to these phase offsets in nonlinear oscillators is the strong frequency pulling~\cite{Slavin2009a,Pandey2008FrequencySystem} that distinctively shifts the bare OMO frequency for each harmonic. Indeed, we can notice in the injection maps of \Cref{fig:2}(h-k) that the locking frequency loci are not symmetric relative to the OMO frequency. For example, \Cref{fig:2}(o) is shifted towards lower frequencies, while \Cref{fig:2}(m) shifts toward higher frequencies. Such shifts are also anticipated by our semi-analytical model and can be traced back to the effective perturbation strength and frequency mismatch in the Adler's model (see ``Methods''). These nonlinearities also highlight the weakness of neglecting the amplitude-phase coupling in the PN modelling of OMOs. 

Another feature that supports the amplitude-phase coupling effects in the PN spectrum, which is not readily captured by the simple model leading to \Cref{eq:PN}, is the presence of the sidebands appearing in \Cref{fig:4}(a) between \SI{20}{\kilo\hertz} and \SI{60}{\kilo\hertz}. In contrast to the fixed-frequency satellite peaks at \SI{150}{\kilo\hertz}, which are caused by parametric mixing with a spurious mechanical mode, these peaks are intrinsic to the nonlinear locking dynamics of OMOs. These sidebands were  discussed by Bagheri et al.~\cite{Bagheri2013PhotonicOscillators} and attributed to coupling between phase and amplitude dynamics that are intrinsic to OMOs. Based upon our amplitude-phase model leading to the effective Adler equations (\Cref{eq:adler}), we derive a quantitative model, in similarity to spin-torque oscillators~\cite{Tortarolo2018}, which predicts both frequency splitting and linewidth of these sidebands. Despite the various approximations necessary, the fitted model agrees remarkably well with the experimental data, as shown in blue/red curves in \Cref{fig:4}(c) and \Cref{fig:4}(d).

In the context of higher-order synchronization, the demonstrated phase-noise performance could be explored towards injection-locked superharmonic frequency dividers~\cite{Rategh2003SuperharmonicDividers,Plessas2011}, which  generate radio-frequency signals at a fraction of a higher frequency reference. Despite the low power-consumption advantage of injection-locked dividers, compared to other technologies, such as regenerative and parametric dividers~\cite{Rategh2003SuperharmonicDividers}, they often suffer from a  narrow lock range. While OMOs offer intrinsically narrower lock ranges compared to electronic injection dividers~\cite{Rategh2003SuperharmonicDividers}, the wide Arnold tongues reported in \Cref{fig:3} suggest that a robust OMO frequency division is feasible. Exploring this strong response to higher harmonics, the experimental schematic of \Cref{fig:4}(e) was implemented to perform the demonstration of an optomechanical frequency division. A low-pass RF filter (\SI{48}{\mega\hertz} cutoff, MiniCircuits SLP-50+) rejects the higher-harmonics generated by the injection-locked OMO and delivers an output signal at a fraction of the injected reference, $f_0/N$. 

The measured frequency divided signals for 2:1, 3:1, and 4:1 locking for a RF power of 0 dBm are shown in \Cref{fig:4}(f). The worst PN performance, obtained in the divide-by-4 case, is better than -70 dBc/Hz and can be significantly improved at higher RF powers, as shown in the red-tone traces in \Cref{fig:4}(b).  Further improvement in phase-noise could be achieved by using devices with higher mechanical quality factor and stronger optical driving power, for instance, double-disk optomechanical devices with mechanical quality factors exceeding $10^4$ and driven at larger amplitudes (using higher optical power) could exhibit a further   PN reduction of 30 dB (see Supplementary Note 3).  These results show that OMO-based frequency dividers can be readily derived from the observed higher-order synchronization. Although there is room for improvement in optomechanical frequency dividers, its ability to generate frequency references in the optical domain could be explored in experiments requiring optical synchronization, such as radio antenna telescopes~\cite{Maleki:2011aa},  optical frequency combs~\cite{Jang2019ObservationCombs}, or coherently linking arrays of optomechanical oscillators with distinct frequencies~\cite{Zhang2015SynchronizationLight}. Given the current state-of-the-art in hybrid integration~\cite{Stern:2018aa} and electro-optical conversion in photonic circuits~\cite{Luan2014}, the demonstrated divider could still ensure the low power consumption expected for injection locking frequency division.

We have experimentally demonstrated an optomechanical oscillator entrained by high-order harmonics and its application as a purely optomechanical frequency division. The wider locking range observed for the higher harmonics, and its theoretical mapping to each nonlinear term in the oscillator dynamics, open new routes to control nonlinear synchronization phenomena in optomechanical oscillators, including the tailoring of the nonlinear response through the laser-cavity detuning and frequency synthesizers optomechanical devices. Furthermore, the importance of nonlinear parametric effects could also significantly impact phase-sensitive amplification~\cite{PhysRevA.102.023507} and  nonlinear sensing~\cite{Brawley2016NonlinearMotion} with optomechanical devices. The demonstrated entrainment should also enable novel configurations for coupling and controlling optomechanical arrays based on dissimilar resonators. The demonstration of locking at fractional harmonics could also be a starting point for further nonlinear dynamics investigations within an optomechanical platform.

\section{Methods}


\noindent \textbf{Optical energy}. The optical energy dependence on the laser-cavity detuning and mechanical displacement is given by,

\begin{equation}
    |a|^2= \frac{\kappa_e}{(\Delta-G x)^2+\kappa^2/4}P_{\text{in}},
    \label{eq:optical_energy}
\end{equation}

\noindent in which two key parameters that will enable the tuning of the OMO nonlinear response arise: the input laser power, $P_{\text{in}}$, and the bare optical detuning, $\Delta=\omega_l-\omega_0$, between the pump laser ($\omega_l)$ and optical mode ($\omega_0$) frequencies; $x$ is the mechanical mode amplitude, $G=\partial\omega/\partial x$ is the optomechanical pulling-parameter, $\kappa$ is the optical mode linewidth and $\kappa_e$ is the external coupling to the bus waveguide \cite{Aspelmeyer2014}.


\noindent \textbf{Effective Adler model}. By employing the Krylov-Bogoliubov-Mitropolsky (KBM) time-averaging method~\cite{bogoliubov1961asymptotic} at the mechanical oscillator equation, an effective Adler's equation may be derived (details in the Supplementary Note 6),

\begin{equation}\label{eq:adler}
    \dot{\Phi} = \nu(\rho) + \varepsilon \frac{\Delta\Omega(\rho)}{2}\sin{\left(\rho\Phi\right)}.
\end{equation}

\noindent where $\Phi$ is the mechanical oscillator phase correction and $\dot{\Phi}$ denote its time derivative; $\nu(\rho)$ is the mean correction of $\Omega_0$ and $\Delta\Omega(\rho)$ is the size of the synchronization window at a particular harmonic $\rho = p/q$. Although many approximations must be carried on, this analysis relates the Taylor series coefficients in \Cref{eq:force_x} with the coefficients $\nu(\rho)$ and $\Delta\Omega(\rho)$ in the effective Adler's model \Cref{eq:adler}, providing a quantitative description of the width hierarchy among the measured ATs.


\noindent \textbf{Experimental setup}. A full schematic of the experimental setup is shown in the Supplementary Note 1, along with optical and mechanical characterization of the bare resonator data. The optical transmission and the RF spectral measurements for the bare resonator properties were taken at low pump powers ($<$ \SI{50}{\micro\watt}). The laser wavelength and detuning are accurately monitored using a Mach-Zehnder Interferometer (MZI) and a HCN gas cell. The cavity is inside a vacuum chamber with pressure of approximately 0.1 mbar and at room temperature. Finally, the transduced signal goes to two detectors: a power meter (PM) that will track the optical mode and a fast photodetector (NewFocus 1617AC Balanced Photodetector) with 800-MHz bandwidth whose electrical output feeds both the electric-spectrum analyzer (ESA, Keysight N9030A) and oscilloscope (OSC, DSO9254A). The phase-noise measurements were performed in the spectral domain using the ESA N9030A phase-noise measurement application (N9068A). There was also a feedback loop between the PM and the TL to lock the signal, preventing the optical resonance to drift due to unwanted external perturbations.

\noindent \textbf{Phase noise}.  To derive the approximate expression for the phase noise (\Cref{eq:PN}), we must start from the general PN expression~\cite{Plessas2011,Zhang1992AOscillators},

\begin{equation}
    \mathcal{L}_{\text{out}}(\Omega)=\frac{(\Delta\Omega_n/n)^2\mathcal{L}_\text{inj}(\Omega)\cos^2\theta+\Omega^2 \mathcal{L}_\text{free}(\Omega)}{\Delta\Omega_n^2\cos^2\theta+\Omega^2}.
    \label{eq:PN_full}
\end{equation}

Since the injection-locking signal is derived from a stable RF frequency source (Agilent PSG E8251), $\mathcal{L}_\text{inj}(\Omega)$, the injection signal PN spectra is orders of magnitude smaller than $\mathcal{L}_\text{free}(\Omega)$, and then $\mathcal{L}_\text{inj}(\Omega)/\mathcal{L}_\text{free}(\Omega) \rightarrow 0$ results in \Cref{eq:PN}. The modulation depth as function of the RF power is given by $\varepsilon=\pi\sqrt{P_{\text{RF}}R}/V_{\pi}$, where $R=$ \SI{50}{\ohm} and $V_{\pi}=$ \SI{5.5}{\volt} is the optical modulator parameter. The phase-angle is given by $\theta=\arcsin\left[(\Omega_0-\Omega_d/n)/\Delta\Omega_n\right]$. A more detailed analysis is given in the Supplementary Note 3 where we show the measured phase noise as a function of the RF power for all the harmonics.

\noindent  \textbf{Simulations}. The acquired data was compared with numerical simulations using Julia language together with well-known and powerful packages like DifferentialEquations.jl, DSP.jl and Sundials.jl. As we are dealing with a stiff system, i.e., there is more than one relevant natural time scale which differ by many orders of magnitude, solvers available in Julia offer a better performance. We simulate the system for a range of modulation depths $\varepsilon$ while the RF signal sweeps around a set of chosen $p:q$ region, revealing the nature of synchronization. With the obtained time trace, we then locally Fourier transformed the data to construct the spectrogram. A detailed discussion on the numerical simulation is available at the Supplementary Note 4. The mechanical mode effective mass and the zero point fluctuation were obtained from COMSOL Multiphysics finite element simulations, $m_{\text{eff}} = 101.82$ pg, $x_{\text{zpf}} =$ \SI{1.536}{\femto\meter}, leading to an optomechanical pulling parameters $G/2\pi=(g_0/2\pi)/x_\text{zpf}=$ \SI{10.546}{\giga\hertz/\nano\meter}.

\noindent \textbf{Data availability}. Further data supporting the findings of this study are openly available at Zenodo at  \href{http://doi.org/10.5281/zenodo.4737381}{DOI:10.5281/zenodo.4737381} upon publication.\\

\section*{Acknowledgements}

This work was supported by São Paulo Research Foundation (FAPESP) through grants 2019/14377-5, 2018/15577-5, 2018/15580-6, 2018/25339-4, 2017/24845-0, 2020/06348-2, 2019/09738-9, Coordenação de Aperfeiçoamento de Pessoal de Nível Superior - Brasil (CAPES)(Financial Code 001). This work was performed in part at the Cornell NanoScale Science and Technology Facility, which is supported by the NSF, its users, and Cornell University.

\section*{Author contributions}

C.C.R. and G.S.W. designed the experiment; C.C.R. performed measurements and data analysis with help from C.M.K. and A.G.P.; C.C.R., C.M.K. and A.G.P. contributed to the theoretical framework. G.S.W. and M.L. designed and fabricated the device; T.P.M.A. and G.S.W supervised the project. All authors contributed to the discussions and preparation of the manuscript.

\section*{Competing interests}

The authors declare no competing interests.

\bibliography{main_refs.bib}

\end{document}


\renewcommand\thesection{Supplementary Note \arabic{section}}

\renewcommand\thefigure{Supplementary Figure \arabic{figure}}
\renewcommand{\figurename}{}

\renewcommand\thetable{Supplementary Table \arabic{table}}
\renewcommand{\tablename}{}

\preprint{APS/123-QED}

\title{Supplementary Information:\\ Optomechanical Synchronization across Multi-Octave Frequency Spans}

\author{Caique C. Rodrigues$^{1,2}$}
\email{ccr@ifi.unicamp.br}
\affiliation{\vspace{0.25cm} $^1$Applied Physics Department, Gleb Wataghin Physics Institute, University of Campinas, Campinas, SP, Brazil \\
$^2$Photonics Research Center, University of Campinas, Campinas, SP, Brazil\\
$^3$Department of Electrical Engineering, Columbia University, New York, New York 10027, USA\\
$^4$Department of Applied Physics and Applied Mathematics, Columbia University, New York, New York 10027, USA}%

\author{Cauê M. Kersul$^{1,2}$}
\affiliation{\vspace{0.25cm} $^1$Applied Physics Department, Gleb Wataghin Physics Institute, University of Campinas, Campinas, SP, Brazil \\
$^2$Photonics Research Center, University of Campinas, Campinas, SP, Brazil\\
$^3$Department of Electrical Engineering, Columbia University, New York, New York 10027, USA\\
$^4$Department of Applied Physics and Applied Mathematics, Columbia University, New York, New York 10027, USA}%

\author{André G. Primo$^{1,2}$}
\affiliation{\vspace{0.25cm} $^1$Applied Physics Department, Gleb Wataghin Physics Institute, University of Campinas, Campinas, SP, Brazil \\
$^2$Photonics Research Center, University of Campinas, Campinas, SP, Brazil\\
$^3$Department of Electrical Engineering, Columbia University, New York, New York 10027, USA\\
$^4$Department of Applied Physics and Applied Mathematics, Columbia University, New York, New York 10027, USA}%

\author{Michal Lipson$^{3,4}$}
\affiliation{\vspace{0.25cm} $^1$Applied Physics Department, Gleb Wataghin Physics Institute, University of Campinas, Campinas, SP, Brazil \\
$^2$Photonics Research Center, University of Campinas, Campinas, SP, Brazil\\
$^3$Department of Electrical Engineering, Columbia University, New York, New York 10027, USA\\
$^4$Department of Applied Physics and Applied Mathematics, Columbia University, New York, New York 10027, USA}%

\author{Thiago P. M. Alegre$^{1,2}$}
\affiliation{\vspace{0.25cm} $^1$Applied Physics Department, Gleb Wataghin Physics Institute, University of Campinas, Campinas, SP, Brazil \\
$^2$Photonics Research Center, University of Campinas, Campinas, SP, Brazil\\
$^3$Department of Electrical Engineering, Columbia University, New York, New York 10027, USA\\
$^4$Department of Applied Physics and Applied Mathematics, Columbia University, New York, New York 10027, USA}%

\author{Gustavo S. Wiederhecker$^{1,2}$}
\email{gsw@unicamp.br}
\affiliation{\vspace{0.25cm} $^1$Applied Physics Department, Gleb Wataghin Physics Institute, University of Campinas, Campinas, SP, Brazil \\
$^2$Photonics Research Center, University of Campinas, Campinas, SP, Brazil\\
$^3$Department of Electrical Engineering, Columbia University, New York, New York 10027, USA\\
$^4$Department of Applied Physics and Applied Mathematics, Columbia University, New York, New York 10027, USA}%

\maketitle


\section{}

\begin{figure}[ht]
    \centering
    \includegraphics[width=\linewidth]{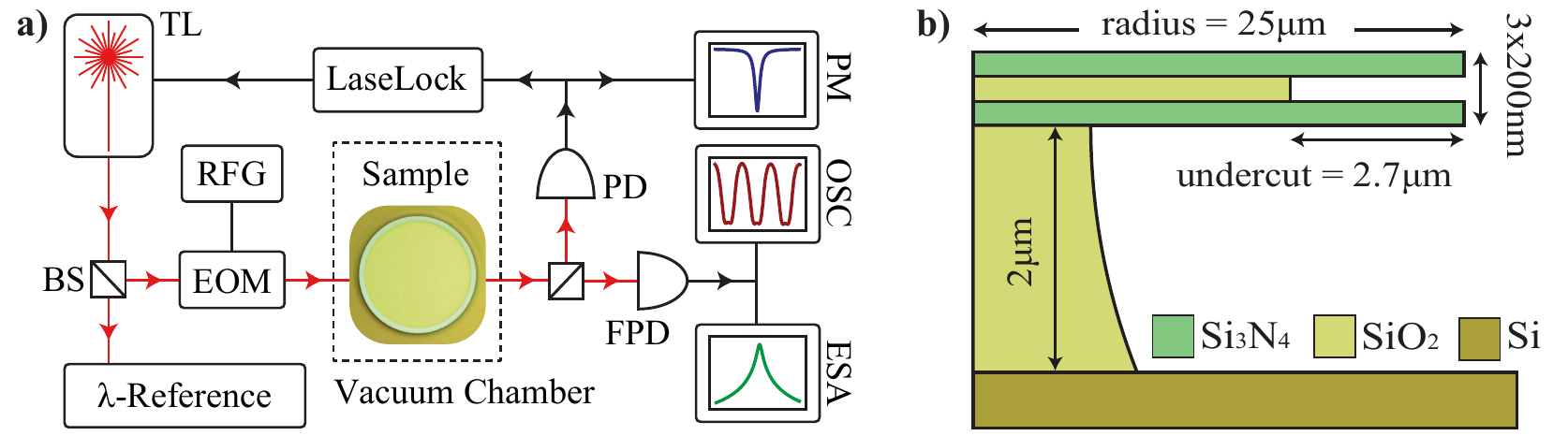}
    \caption{\textbf{a)} Experimental setup used in the article. A tunable laser (TL) goes into a beam splitter (BS), in which one of the arms goes to a HCN cell wavelength reference, and the other arm goes to a electro-optical modulator (EOM) that is controlled by a radio frequency generator (RFG, Agilent PSG E8251). The modulated field after interacting with the sample inside a vacuum chamber of $\approx$ 0.1 mbar goes to another beam splitter which we finally obtain our results. The output signal then reaches a fast photodetector (FPD), which measures both the temporal trace using an oscilloscope (OSC, DSO9254A) and also the spectral content at a electrical spectrum analyzer (ESA, Keysight N9030), but also a slow photodetector (PD) which gives the Lorentzian shape optical transmission. The final part of the setup is a feedback loop (LaseLock) that goes back into the tunable laser that makes the laser wavelength stable by self referenciation, avoiding unwanted drifts during the data acquisition; \textbf{b)} Illustration of the nitride double disk cavity geometry used in the experiment.}
    \label{fig:1}
\end{figure}

\textbf{Cavity Characterization.} The whole experimental setup and the cavity geometry are shown in \ref{fig:1}.

\vspace{0.0cm}

\begin{figure}[ht]
    \centering
    \includegraphics[width=\linewidth]{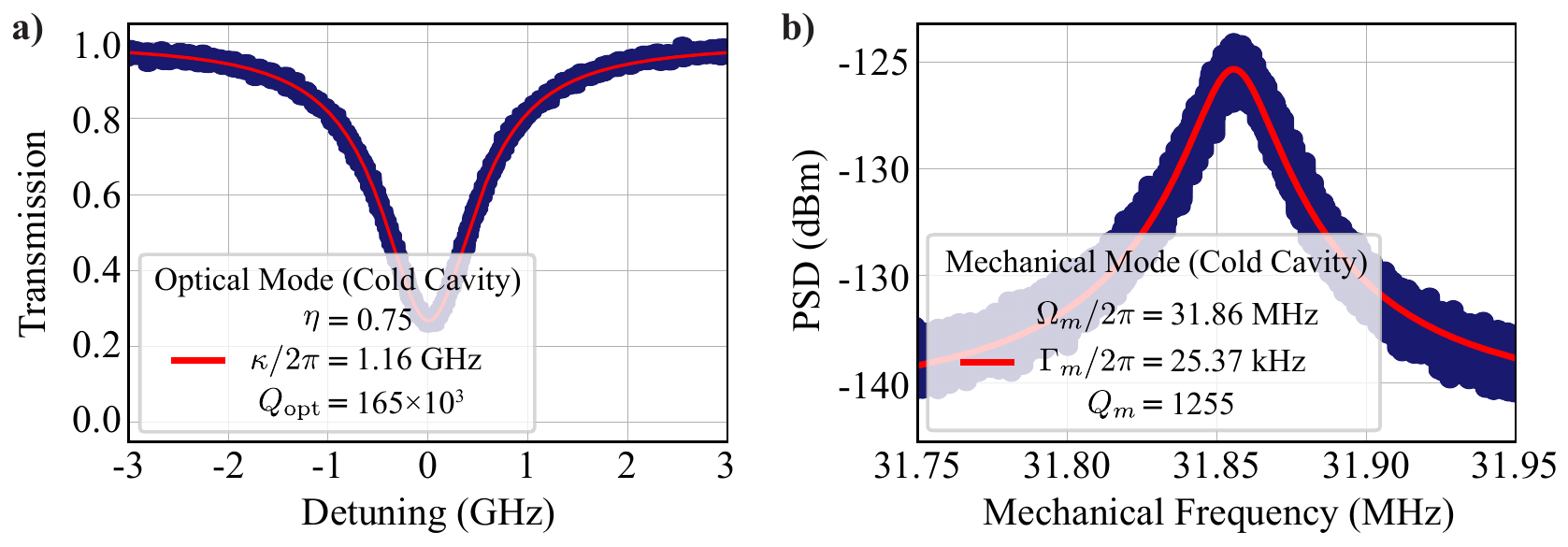}
    \caption{\textbf{a)} Experimental optical transmission spectrum of the cavity; \textbf{b)} Experimental power spectrum density (PSD). The best fits of both curves are shown in red.}
    \label{fig:2}
\end{figure}

The optical and the mechanical modes used in this experiment are shown in \ref{fig:2}, with their best fits in red. The model of these curves are given by well-known \Cref{eq:fit_opt} and \Cref{eq:fit_mech}

\begin{equation}\label{eq:fit_opt}
    T(\Delta) = \left|\frac{s_{\text{out}}}{s_{\text{in}}}\right|^2 = \frac{\left(1 - 2\eta\right)^2 + \frac{4\Delta^2}{\kappa^2}}{1 + \frac{4\Delta^2}{\kappa^2}} \quad \quad \quad \quad \text{(Optical Transmission Spectrum)}
\end{equation}

\begin{equation}\label{eq:fit_mech}
    \mathcal{S}_{PP}[\Omega] = \mathcal{S}^{\text{min}}_{PP} + \frac{\left(\mathcal{S}^{\text{max}}_{PP} - \mathcal{S}^{\text{min}}_{PP}\right)\left(\Gamma_m\Omega_m\right)^2}{\left(\Omega^2 - \Omega_m^2\right)^2 + \left(\Gamma_m \Omega\right)^2} \quad \quad \quad \quad \text{(Power Spectral Density)}
\end{equation}

\vspace{0.2cm}

The value measured for the vacuum optomechanical coupling rate was $g_0/2\pi = 16.2$ kHz, where we have followed the M. L. Gorodetksy et al. article \cite{Gorodetksy:10}. The function $s_{\text{in}}^2$ can be interpreted as the power reaching the cavity, i.e., $s_{\text{in}}^2 = P_{\text{in}}$, which is also valid for the output field $s_{\text{out}}^2 = P_{\text{out}}$. The power spectral density $\mathcal{S}_{PP}$ (or just PSD) shown in \Cref{eq:fit_mech} is in dBm units. The parameter $\eta = \kappa_e/\kappa$ is the coupling between the optical fiber taper and the cavity. For now on we are omitting the sub-index of $s_{\text{in}}(t) \rightarrow s(t)$, because we are not using $s_{\text{out}}$ in any future calculations, so there will be no ambiguity in just writing $s(t)$ for the input field.


\section{}

\begin{figure}[ht]
    \centering
    \includegraphics[width=\linewidth]{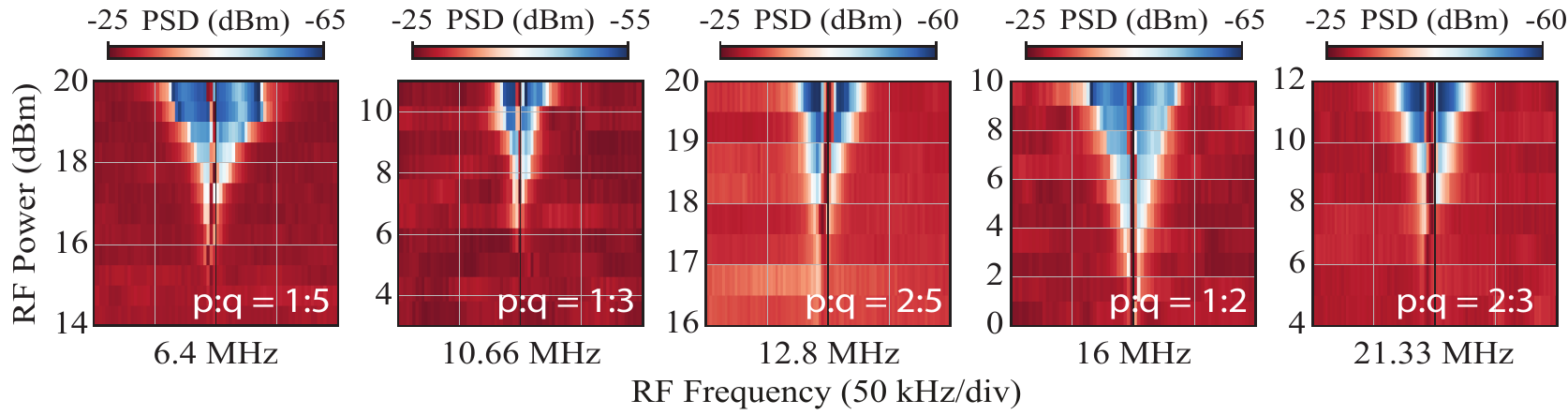}
    \caption{Experimental fractional order Arnold tongues maps. The $p:q$ order is shown in white inside each map, which are arranged from lower frequency to higher frequencies from the left to the right.}
    \label{fig:3}
\end{figure}

\textbf{Fractional Synchronization.} As mentioned in the article we also observed several fractional order synchronizations, i.e., $\rho=p/q$ not an integer. We have shown in the main article, however, only the threshold to observe the tip of the Arnold tongues; here, we present in \ref{fig:3} the whole experimental map obtained.
 
These maps, however, requires a really strong modulation depth $\varepsilon$, impossible to study them in the weak perturbation regime of our semi-analytical model, as we are drastically changing the dynamic of the system. The importance of these data is to prove the existence of this kind of injection locking in optomechanics and to motivate the study of such phenomenon in future works, maybe the possibility of achieving such regimes using weak perturbations with different experimental parameters or even in some other cavity design to enhance these effects.

\newpage

\section{}

\begin{figure}[ht]
    \centering
    \includegraphics[width=\linewidth]{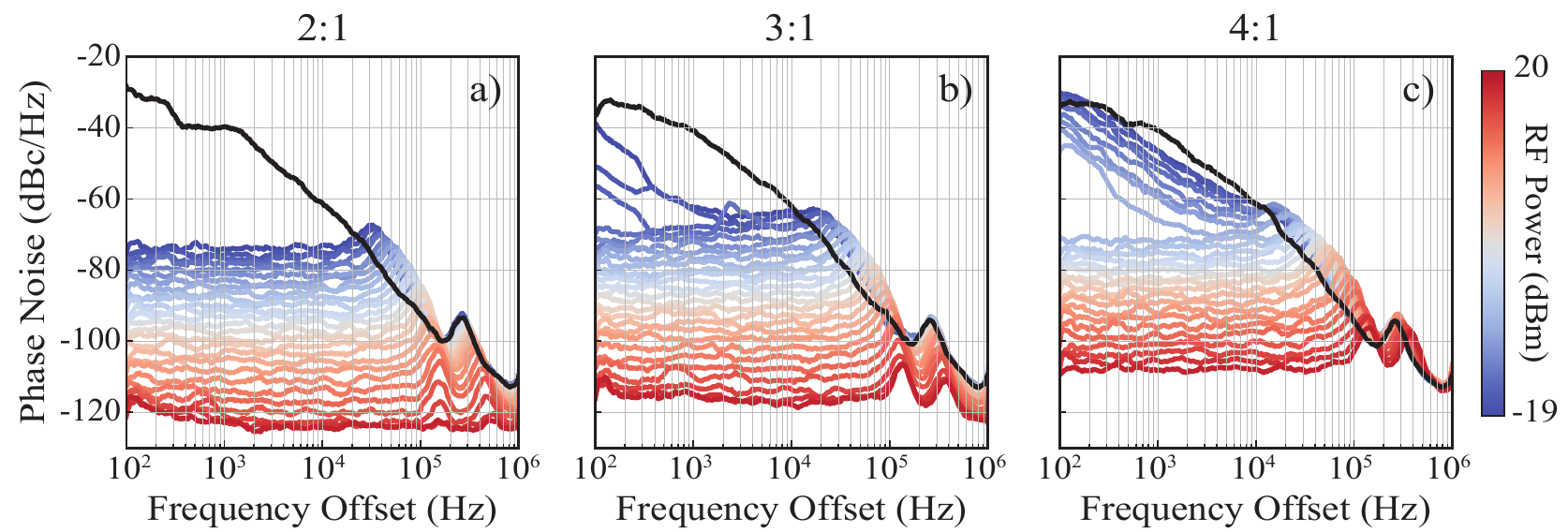}
    \caption{Phase noise spectral densities for the cases $2:1$, at \textbf{a)}, $3:1$, at \textbf{b)}, and $4:1$, \textbf{c)}, for various modulation depth. The black curve is the OMO free-running PN for reference.}
    \label{fig:13}
\end{figure}

\textbf{Phase Noise Analysis and Routes to Frequency Division Optimization.} In the main text we have shown only the phase noise measurements for the $4:1$ injection, here we will show the PN density spectra for the high-harmonics, 2 : 1 to 4 : 1, as a function of the RF power, as shown in \ref{fig:13}. For the smallest modulation depth (-19 dBm $\approx 1.5\%$) the $2:1$ PN is flat around -70 dBc/Hz, in contrast with the $3:1$ and $4:1$ cases. For small modulations, both the $3:1$ and $4:1$ PN spectra appear to be transitioning from the OMO free-running spectrum to the injection-locked regime characterized by the flat plateau. This is expected as the farther we are from $\Omega_0$ because smaller the interaction, in such a way that higher modulation depths are needed to achieve the same low PN levels.

\begin{figure}[ht]
    \centering
    \includegraphics[width=\linewidth]{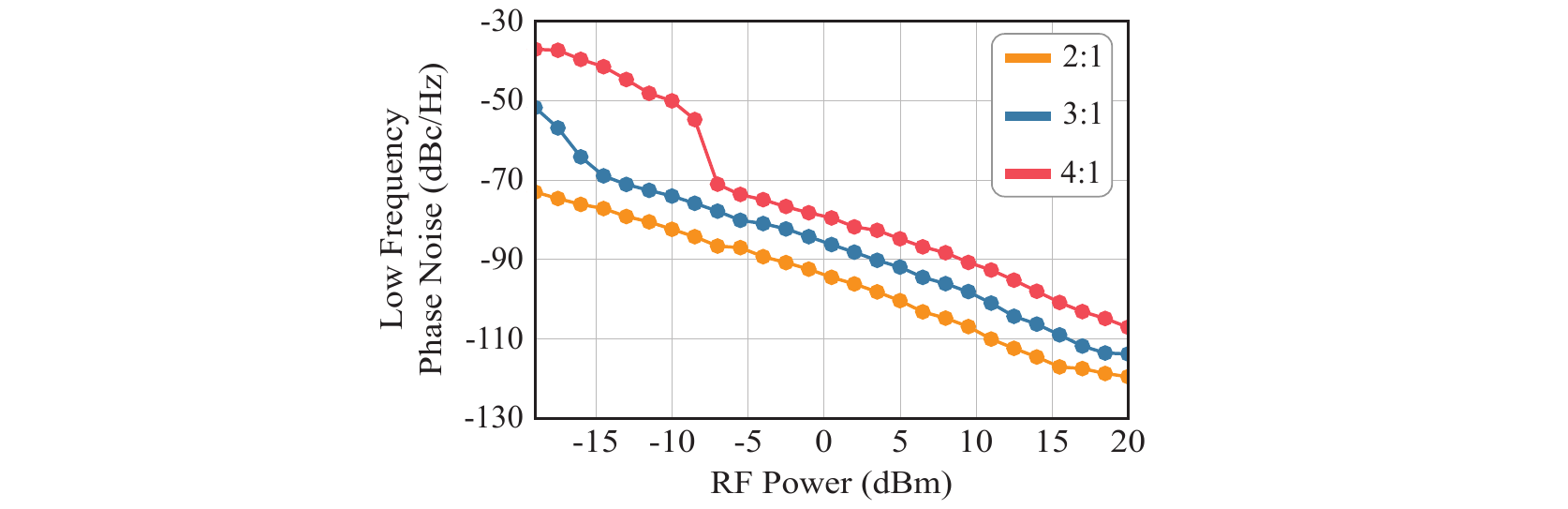}
    \caption{Average phase noise around 100 Hz and 1 kHz for high harmonic injections as function of the RF power.}
    \label{fig:16}
\end{figure}

A compact way to visualize the evolution of the phase noise as a function of the RF power is averaging the phase noise around low frequencies, as done in \ref{fig:16}.

We address possible paths to optimize the injection locked phase noise. The model used for superharmonic injection phase noise is

\begin{equation}
    \mathcal{L}_{\text{out}}(\Omega)=\frac{(\Delta\Omega_n/n)^2\mathcal{L}_\text{inj}(\Omega)\cos^2\theta+\Omega^2 \mathcal{L}_\text{free}(\Omega)}{\Delta\Omega_n^2\cos^2\theta+\Omega^2},
    \label{eq:PN_full}
\end{equation}

\noindent which is a weighted average of $\mathcal{L}_\text{inj}(\Omega)$ and $\mathcal{L}_\text{free}(\Omega)$. If any one of these is much smaller than the other, i.e., $\mathcal{L}_\text{inj}(\Omega) \ll \mathcal{L}_\text{free}(\Omega)$, we \textit{cannot} expect $\mathcal{L}_{\text{out}}(\Omega) \approx \mathcal{L}_\text{inj}(\Omega)$, which would be desirable to a frequency divider. That being said, we must find ways to improve $\mathcal{L}_\text{free}(\Omega)$ such that $\mathcal{L}_\text{free}(\Omega)$ approaches $\mathcal{L}_\text{inj}(\Omega)$. According to \cite{PhysRevA.90.023825} and \cite{Mathai:19}, an optomechanical cavity in the unresolved sideband regime is dominated by thermomechanical noise and its phase noise spectral density is given by

\vspace{0.0cm}
\begin{equation}\label{leeson}
    \mathcal{L}_\text{free}(\Omega) = \left(\frac{2\Gamma_m}{n_x}\right)\left(\bar{n}_\text{th} + \frac{1}{2}\right)\left(\frac{1}{\Omega^2} + \frac{\nu_{\text{om}}'^{2}}{\Omega^2}\frac{1}{\gamma_{\text{om}}'^{2} + \Omega^2} + \frac{\eta_{I}^2}{\gamma_{\text{om}}'^{2} + \Omega^2}\right),
\end{equation}
\vspace{0.15cm}

\noindent in which three ``general rules" towards phase noise reduction are noticed: reducing $\Gamma_m$ (mechanical oscillator's linewidth); reducing $\bar{n}_\text{th}$ (thermal phonon number); and increasing $n_x$ (coherent phonon number).

\begin{figure}[ht]
    \centering
    \includegraphics[width=\linewidth]{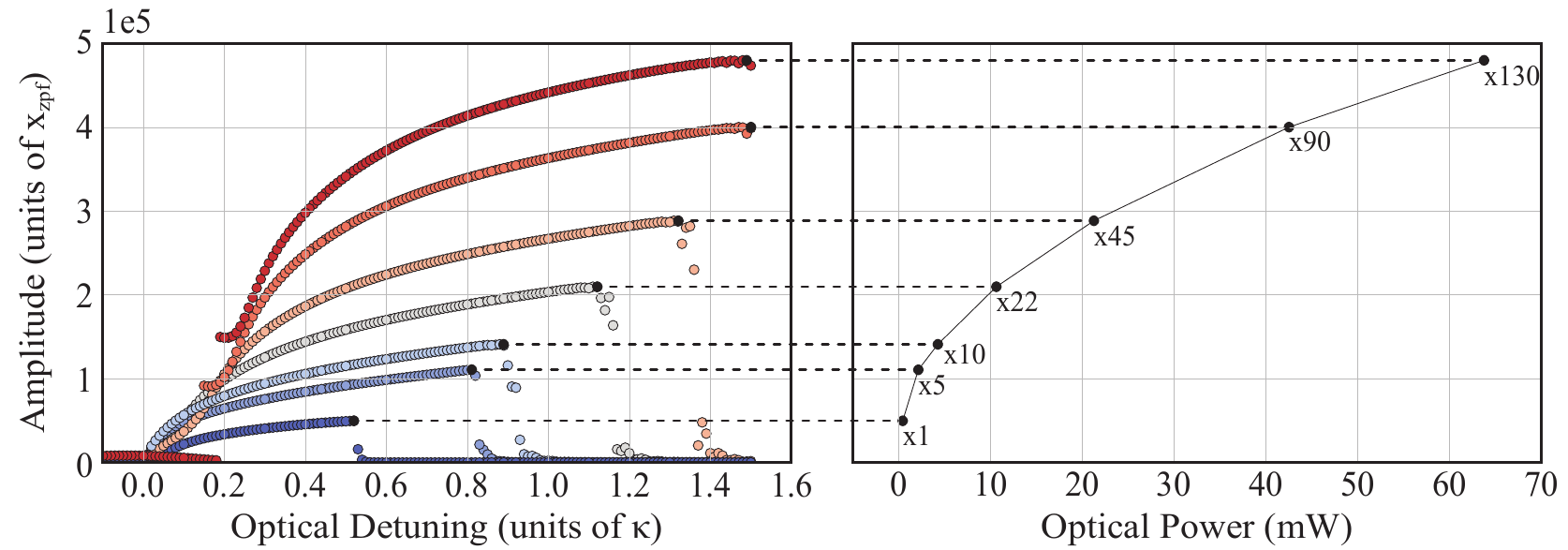}
    \caption{\textbf{a)} Mechanical oscillator's amplitude, in units of $x_{\text{zpf}}$, as a function of the optical detuning $\Delta_x$, in units of $\kappa$; \textbf{b)} Maximum oscillator's amplitude for a given optical power $P_0$, in mW. The floating numbers indicates how many times we must multiply the actual pump power used in the experiment to reach that level. The $\times$1 case is the actual numerical simulation using the data obtained in the laboratory, which is $P_0 = 480\mu$W.}
    \label{fig:15}
\end{figure}

Quality factors up to $Q_m \approx 10000$, 10 times larger than in our experiment, are reported in silicon nitride double disks~\cite{Zhang2014}. Such large mechanical quality factors were obtained just by manipulating the thickness of the nitride film. Also, increasing the optical pump power $P_0$ should readily increase the coherent phonon occupation $n_x$. To obtain an estimate on a feasible amplitude enhancement, we simulate the oscillation amplitude $x$ (in units of its zero point fluctuation, $x_{\text{zpf}}$) as a function of the optical detuning $\Delta_x$, where the subindex indicates that we already counted the static optomechanical shift $x_0$, i.e., $\Delta_x = \omega_l - \omega_0 - Gx_0$. Our simulations were performed for a set of optical powers, as shown in \ref{fig:15}, to find out the precise scaling of $n_x$  with $P_0$. The numerical simulation details and the parameters used are covered in the next section.

The terms $\nu_{\text{om}}'$, $\gamma_{\text{om}}'$ and $\eta_I$ are the frequency shift due to amplitude change at the limit cycle, the damping rate of the amplitude fluctuation and transfers from displacement amplitude noise to photon number phase noise, respectively \cite{PhysRevA.90.023825}. As \Cref{leeson} show us, all these terms also contribute to the phase noise level, however, as we are using an optomechanical cavity in an unresolved sideband regime, we are neglecting their contributions because the term $1/\Omega^2$ dominates. Therefore, combining higher quality factors with larger oscillation amplitudes, we could achieve a net 30 dBc/Hz improvement, paving the way towards future improvements in optomechanical frequency dividers.

\newpage

\section{}

\textbf{Numerical Simulation.} The numerical simulations of this section are not straightforward to perform, and it is worth discussing how they were carried out carefully. One issue faced when solving the following coupled nonlinear ODE

\begin{figure}[ht]
    \centering
    \includegraphics[width=\linewidth]{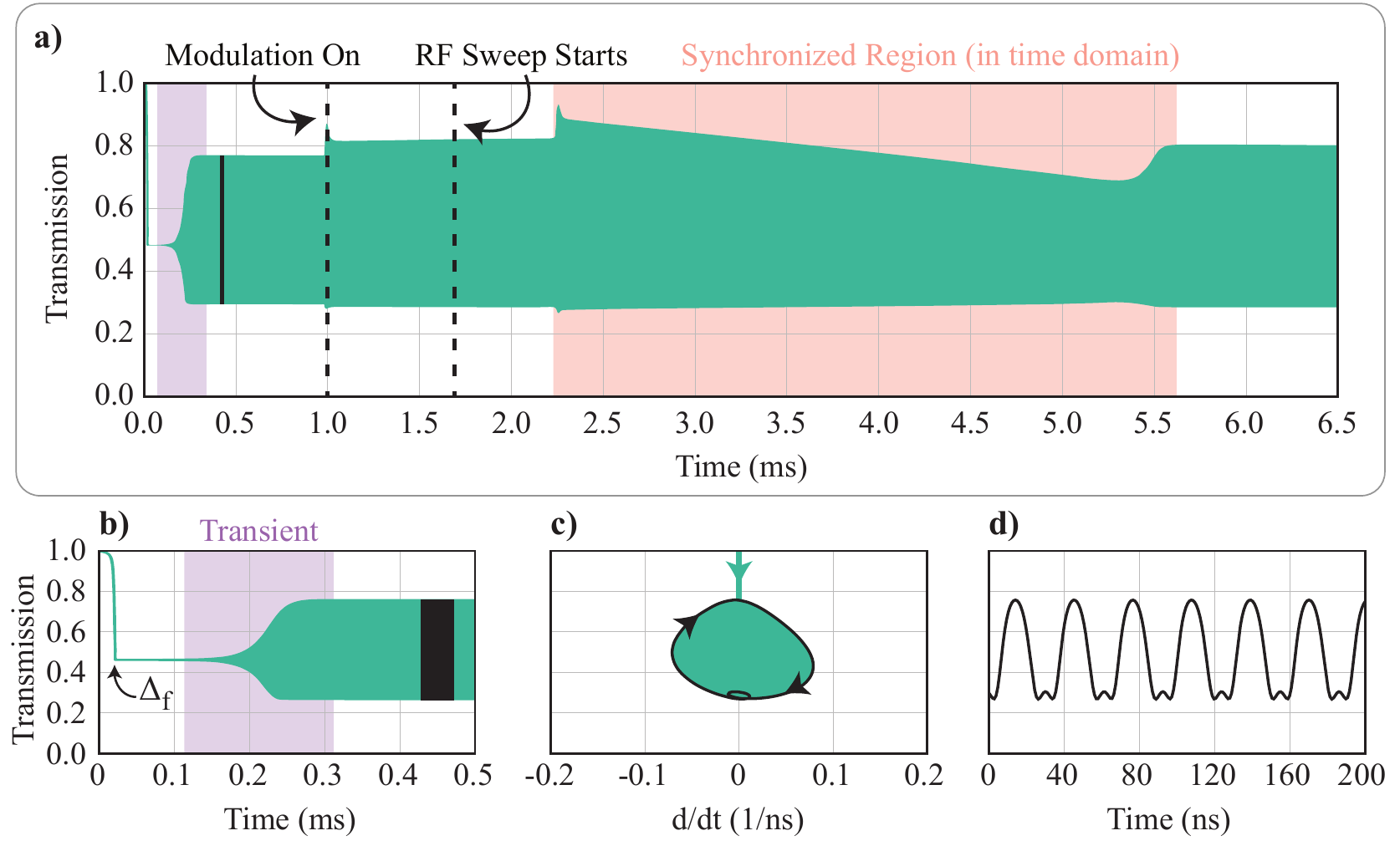}
    \caption{\textbf{a)} Complete time domain simulation, showing important aspects of the synchronization. The purple region is the transient region which the mechanical oscillator gains amplitude. The two vertical dashed black lines shows where exactly we turned on the modulation $\varepsilon$ and the RF sweep $\dot{\Omega}_d$. The pink region is where injection locking is happening; \textbf{b)} Transient region of \ref{fig:4}(a) showing the regime from non-oscillating cavity to self-sustained oscillation; \textbf{c)} Phase space of \ref{fig:4}(b); \textbf{d)} Temporal trace of the black part of \ref{fig:4}(b).}
    \label{fig:4}
\end{figure}

\vspace{-0.3cm}
\begin{equation}\label{eq:optomechanics_1}
    \dot{a} = i\Delta(t) a - \frac{\kappa}{2}a - iGxa + \sqrt{\kappa_e}s_{0}\sqrt{1 + \varepsilon(t)\sin{\Theta_d(t)}} \quad \quad \text{and} \quad \quad \ddot{x} + \Gamma_{m}\dot{x} + \Omega_{m}^{2}x = -\frac{\hslash G}{m_{\text{eff}}}|a|^{2}
\end{equation}

\noindent is the stiff nature of the system, characterized by the need of very small discretization steps despite the relative smoothness of the solutions. To tackle this system, we used well known numerical packages DifferentialEquations.jl, FFTW.jl, Sundials.jl and DSP.jl available in Julia language which implements robust methods for such systems. The simulation was done as follows: we first set an optical detuning function $\Delta(t)$ to sweep linearly from $\Delta_i$ to $\Delta_f$, where the sub-indexes $i$ and $f$ means initial and final, respectively. We have chosen $\Delta_i > 0$ because we want to access the blue side of the optical mode, where the self-sustained dynamic is naturally accessible. After reaching $\Delta_f$, we wait a few cycles of the mechanical oscillator to make sure the system is in a stationary regime and then turned on the modulation depth $\varepsilon(t)$, in which we modeled as a Heaviside step function. With the modulation depth online we, once again, waited a few microseconds to stabilize the energy inside the cavity, and then finally turned on the RF frequency sweep. In the laboratory our RF frequency sweep was linear between $\Omega_{d}^{i}$ and $\Omega_{d}^{f}$ with constant velocity $d\Omega_{d}/dt = \dot{\Omega}_d$, so we modeled $\Theta_d(t)$ as a parabola, i.e., $d\Theta_d(t)/dt = \Omega_{d}(t) = \Omega_{d}^{i} + \dot{\Omega}_d t$. The value chosen for $\dot{\Omega}_d$ need to be small to guarantee adiabaticity, which clearly is the case in the laboratory. A good threshold for adiabaticity is to sweep the RF tone over the mechanical resonance (of linewidth $\Gamma_m$) within the mechanical lifetime, $\tau_{m} \approx 2\pi/\Gamma_m$, i.e., $\dot{\Omega}_d \approx \Gamma_m/\tau_{m} \approx \Gamma_m^2/2\pi$. For our purposes, a RF frequency sweep velocity of $\dot{\Omega}_d \approx 0.1\Gamma_m^2$ was enough to ensure adiabaticity. A summary of all said is shown in \ref{fig:4}, highlighting the main aspects of the dynamic.

\vspace{0.0cm}
\vspace{-0.0cm}

\begin{figure}[ht]
    \centering
    \includegraphics[width=0.9\linewidth]{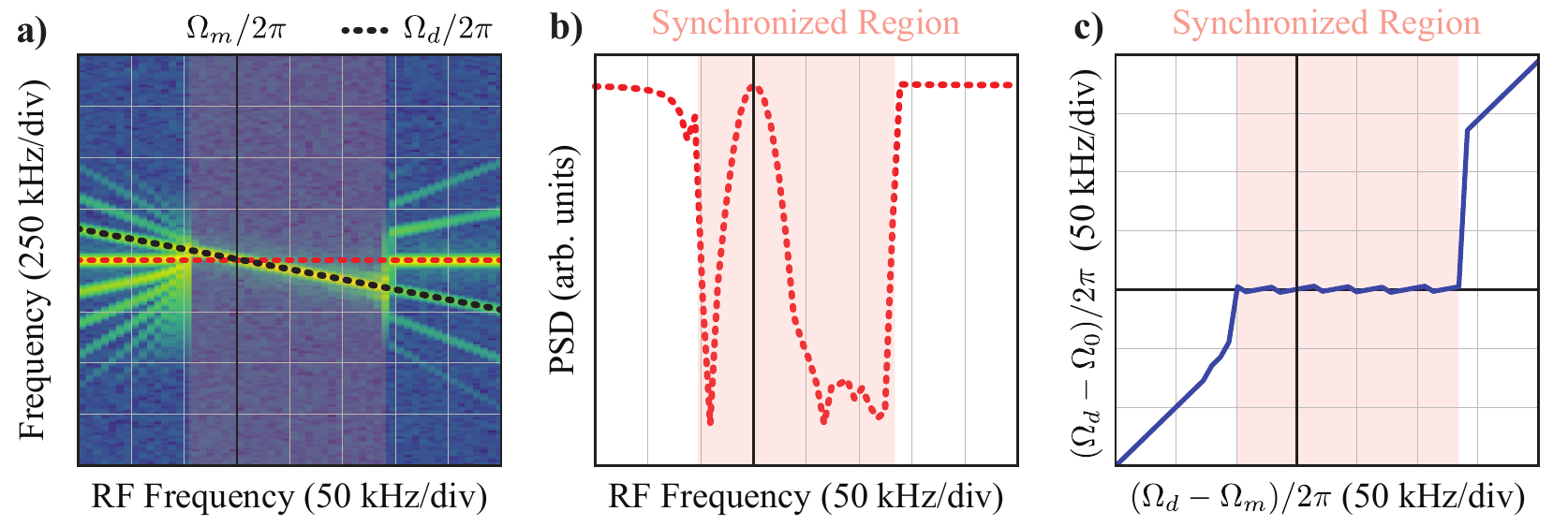}
    \caption{\textbf{a)} Spectrogram of the transmission signal (\ref{fig:4}(a)) after the RF sweep is turned on. The dashed black line is the value of the drive frequency. The vertical bold black line is the value of the mechanical mode frequency; \textbf{b)} Horizontal dashed red slice of the spectrogram shown in \textbf{a)}; \textbf{c)} Typical plot of synchronization systems showing the mismatch from driving frequency and bare oscillation frequency, making clear where these became the same, defining a synchronized state. We have used $\varepsilon = 2\%$ for these simulations.}
    \label{fig:5}
\end{figure}

These are the raw data that we obtain from the simulation. To obtain from these data the Arnold tongues we can take the length of the synchronized region of \ref{fig:4}(a) - the pink region of the plot -  for each modulation depth $\varepsilon$. However, we must clarify how we find this pink region, i.e., the specific point where we say that synchronization occurs is a bit blur in the time domain, and that's why we construct a spectrogram, which is the Fourier transform of our signal in function of time, as shown in \ref{fig:5}(a). However, instead of plotting the spectrogram as a function of time, as we know the value of the driving RF frequency $\Omega_d$ for each time $t$, we can plot the spectrogram already in function of the RF frequency, and that is what was done in \ref{fig:5}(a). One way to obtain the synchronized region is to take the horizontal slice of this spectrogram just above the mechanical oscillation frequency $\Omega_0/2\pi$, which is the horizontal dashed red line, and its plot is shown in \ref{fig:5}(b). A second way, which is more well known in the literature, is to plot the difference between the driving frequency and oscillator's frequency ($\Omega_d - \Omega_0$) as a function of the drive frequency itself (or, in our case, the driving frequency minus a constant natural mechanical frequency $\Omega_m$), as shown in \ref{fig:5}(c).

\begin{figure}[ht]
    \centering
    \includegraphics[width=0.95\linewidth]{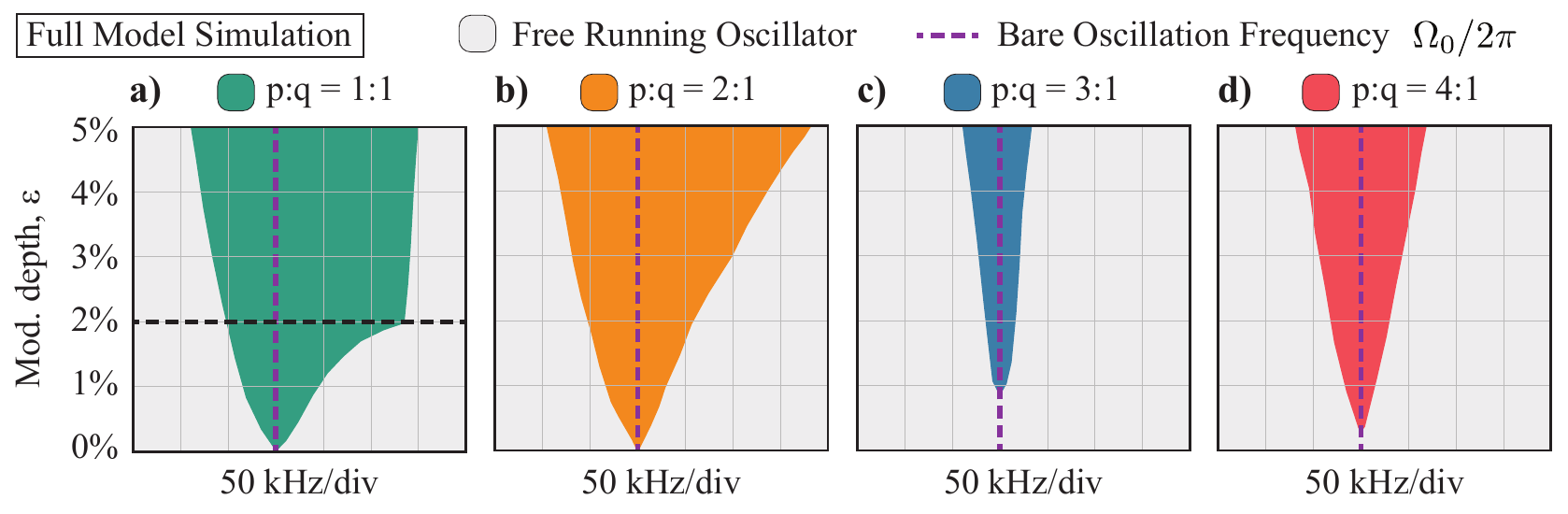}
    \caption{Simulated Arnold tongues using injection frequency $\Omega_d = p\Omega_{0}/q$ for the cases $p = \{1,2,3,4\}$ and $q=1$, in order, from \textbf{a)} to \textbf{d)}. To simulate these maps we used \Cref{eq:optomechanics_1}, being these the ones that we will always call ``full model". The dashed black line at \textbf{a)} is the region that we analyzed in \ref{fig:5}(c).}
    \label{fig:6}
\end{figure}

The Arnold tongues constructed using the explanation above are shown in \ref{fig:6}, which was already presented at the article as Fig. 3(a). As we can see, the simulation shows bigger synchronized region for the case $p:q = 2:1$ than $1:1$, and also a pretty wide $4:1$ AT, but a small $3:1$, the same trend of the experimental data. \ref{table:1} shows the parameters used in the simulations done and \ref{fig:7} shows the conversion from RF power, in dBm, to modulation depth $\varepsilon$, in $\%$, which is based in experimental data. The actual formula for the RF power $P_{RF}$ is shown inside the plot.

\vspace{-0.0cm}
\vspace{0.1cm}

\newpage

\begin{table}[ht]
    \begin{minipage}[ht]{0.4\linewidth}
        \centering
        \begin{tabular}{|c|c|}
            \hline
            \textbf{Parameters} & \textbf{Values}              \\ \hline
            $P_{0}$                & 425 $\mu$W                       \\ \hline
            $\lambda$              & 1560nm                       \\ \hline
            $\Delta_i$              & 8$\kappa$                       \\ \hline
            $\Delta_f$              & 0.35$\kappa$                    \\ \hline
            $d\Delta/dt$            & 10$^2 \Gamma_m^2$ \\ \hline
            $d\Omega_d/dt$           & 0.075$\Gamma_m^2$                 \\ \hline
            $\eta$                 & 0.75                         \\ \hline
            $\kappa/2\pi$            & 1.16 GHz                     \\ \hline
            $Q_{\text{opt}}$                & 165000                       \\ \hline
            $\Omega_m/2\pi$           & 31.86 MHz                    \\ \hline
            $\Gamma_m/2\pi$           & 25.37 kHz                    \\ \hline
            $Q_m$                  & 1255                         \\ \hline
            $g_0/2\pi$               & 16.2 kHz                     \\ \hline
        \end{tabular}
        \caption{Parameters values used in every simulations, unless explicitly mentioned the opposite.}
        \label{table:1}
    \end{minipage}\hfill
    \begin{minipage}[ht]{0.56\linewidth}
        \centering
        \includegraphics[width=0.8\linewidth]{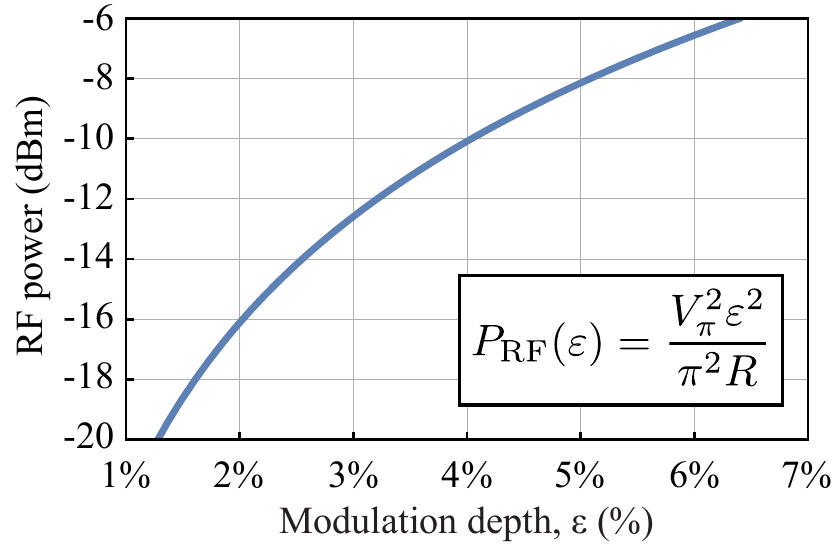}
        \captionof{figure}{Conversion from modulation depth, in percentage, to RF power, in dBm. The inset shows the actual formula, in S.I. units, of this graph. $V_\pi =$ 5.5 V and $R =$ 50 $\Omega$.}
        \label{fig:7}
    \end{minipage}
\end{table}

\section{}

\textbf{Semi-Analytical Model.} The Hamiltonian of our system, regardless dissipative considerations, can be modeled by

\vspace{-0.35cm}
\begin{equation}\label{eq:hamiltonian}
    H = \hslash\left(\omega_{0} + Gx\right)a^{\dagger}a + \frac{p^2}{2m_{\text{eff}}} + \frac{m_{\text{eff}}\Omega_m x^2}{2} + i\hslash\sqrt{\kappa_e}s_0\left(e^{-i\omega_l t} a^{\dagger} - e^{i\omega_l t}a\right) ,
\end{equation}

\noindent in which $\hslash$ is the reduced Planck's constant, $\omega_{0}$ is the unperturbed angular frequency of the optical mode, $a^{\dagger}$ and $a$ are the creation and annihilation operators for photons with energy $\hslash\omega_{0}$, respectively, $G$ is the first order coefficient of the Taylor expansion of $\omega(x) = \omega_0 + Gx$ evaluated at mechanical equilibrium position $\left(\text{i.e., } G = (d\omega/dx)|_{x = 0}\right)$, $p$ and $x$ are the momentum and position operator of the mechanical oscillator, respectively, $\Omega_{m}$ is the unperturbed angular frequency of the mechanical mode, $m_{\text{eff}}$ is the mechanical oscillator effective mass, $i$ is the complex unity, $\kappa_e$ is the external optical coupling rate, $s_{0}^2$ is the input power and $\omega_l$ is the optical pump angular frequency. As long as we are not interested in quantum phenomena we can study our dynamical system just looking to the average value of these operators, and we can also introduce the optical and mechanical loses $\kappa$ and $\Gamma_m$ directly in the equations of motion \cite{Aspelmeyer2014} as

\vspace{-0.4cm}
\begin{equation}\label{eq:optomechanics_2}
    \dot{a} = i\Delta a - \frac{\kappa}{2}a - iGxa + \sqrt{\kappa_e}s_{0} \quad \quad \text{and} \quad \quad \ddot{x} + \Gamma_{m}\dot{x} + \Omega_{m}^{2}x = -\frac{\hslash G}{m_{\text{eff}}}|a|^{2},
\end{equation}

\noindent where we already changed $a$ to the slow rotating frame of reference $a \rightarrow ae^{-i\omega_l t}$ to let the equation autonomous. We define the bare optical detuning $\Delta = \omega_l - \omega_{0}$ as the difference between optical pump frequency and unperturbed optical mode frequency. To introduce the amplitude modulation used in the experiment we can simply multiply $s_{0}$ by a factor $\sqrt{1 + \varepsilon\sin{\Theta_d(t)}}$ in \Cref{eq:optomechanics_2}, i.e., $s(t) = s_0\sqrt{1 + \varepsilon\sin{\Theta_d(t)}}$, in which $\varepsilon$ is the modulation depth and is correlated with the RF power as shown in \ref{fig:7}. The term $\Theta_d(t)$ is the phase of this modulation, in which most of the time will just be $\Omega_d t$. Simulating \Cref{eq:optomechanics_2} as it is shown require us to know $G$ and $m_{\text{eff}}$ but, because they are normalization dependent, we are going to avoid this using $g_0 = Gx_{\text{zpf}}$, the optomechanical single-photon coupling strength, and also the $x_{\text{zpf}} = \sqrt{\hslash/2m_{\text{eff}}\Omega_m}$, the zero points fluctuation amplitude of $x$. New normalizations will be used to study self-sustained oscillations, which are given by

\vspace{-0.2cm}
\begin{equation}\label{eq:normalization_1}
    x(t) = x_0 + \delta x(t) = \left(\frac{\kappa}{2g_0}\right)\sqrt{\frac{\hbar}{2m_{\text{eff}}\Omega_m}}\left(\widetilde{x}_{0}+\widetilde{x}(t)\right) \quad \quad \text{and} \quad \quad t = \frac{\widetilde{t}}{\Omega_m},
\end{equation}
\vspace{-0.1cm}

\noindent where all tilde variables are now adimensional. The terms $x_0$ and $\delta x(t)$ are the DC and AC components of $x(t)$, respectively, being $\widetilde{x}_0$ and $\widetilde{x}(t)$ their adimensional version. We can then rewrite \Cref{eq:optomechanics_2} as

\begin{figure}[ht]
    \begin{minipage}[b]{0.56\linewidth}
        \justify \normalsize{remembering that $\delta\Gamma_m^{\text{linear}}$ is in units of $\Omega_m$. Solving then for $\widetilde{\tau}$}
        
        \vspace{-0.0cm}
        \begin{equation}\label{eq:tau}
            \widetilde{\tau} = \frac{1}{2\left(\frac{2\Delta_x}{\kappa}\right)}\left[\frac{1 + \left(\frac{2\Delta_x}{\kappa}\right)^2}{1+\left(\frac{2\Delta_x}{\kappa} - \frac{2\Omega_m}{\kappa}\right)^2} - \frac{1 + \left(\frac{2\Delta_x}{\kappa}\right)^2}{1+\left(\frac{2\Delta_x}{\kappa} + \frac{2\Omega_m}{\kappa}\right)^2}\right].
        \end{equation}
        \vspace{0.3cm}
    
        We must emphasize here that our $\widetilde{\tau}$ should, rigorously, be rewritten as $\widetilde{\tau}_{\text{linear}}$, because that is just the first order correction of $\widetilde{\tau}$. Nevertheless, the function $\widetilde{\tau} = \widetilde{\tau}(\Delta_x)$ has every property that we expect: (i) is positive for every value of $\Delta_x$, (ii) it is approximately $\Omega_m/\kappa$ and (iii) is also consistent with the fact that far from the resonance there is no mechanical response, i.e., $\widetilde{\tau}(|\Delta_x| \gg \kappa) = 0$, as shown in \ref{fig:8} using our experimental parameters and also for three other cases.
    \end{minipage}\hfill
    \begin{minipage}[b]{0.4\linewidth}
        \centering
        \includegraphics[width=\linewidth]{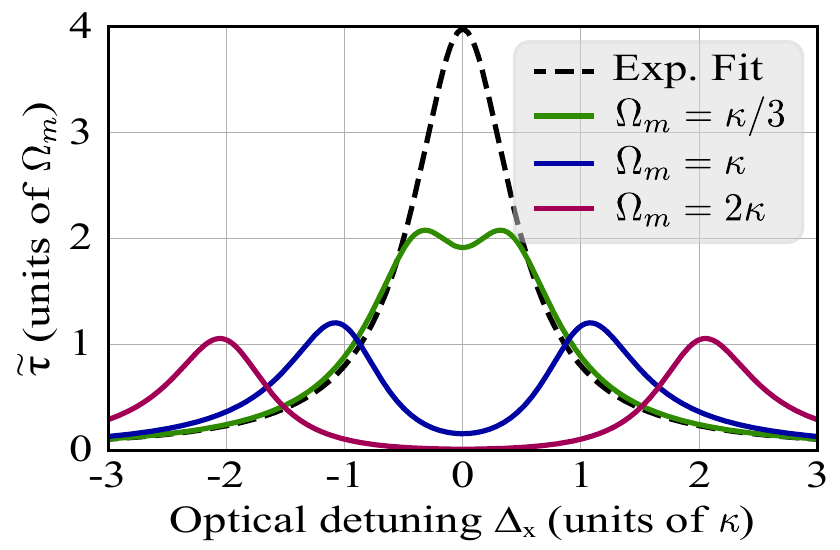}
        \caption{Adimensional linear mechanical relaxation time $\widetilde{\tau}$ (in units of $\Omega_m/\kappa$) as function of the optical detuning $\Delta_x$ (in units of $\kappa$).}
        \label{fig:8}
    \end{minipage}
\end{figure}

\vspace{-0.2cm}
\begin{equation}\label{eq:optomechanics_3}
    \frac{da}{d\widetilde{t}} = -\frac{\kappa}{2\Omega_m}a + i\left(\frac{\Delta_x}{\Omega_m}-\frac{\kappa}{2\Omega_m}\widetilde{x}\right)a + \sqrt{\frac{\kappa_e}{\Omega_m}}\frac{s}{\sqrt{\Omega_m}} \quad \quad \text{and} \quad \quad \frac{d^2\widetilde{x}}{d\widetilde{t}^2} + \frac{1}{Q_m}\frac{d\widetilde{x}}{d\widetilde{t}} + \widetilde{x} = -\widetilde{x}_0 - \frac{\mathcal{C}_0}{Q_m}|a|^{2},
\end{equation}

\noindent where the new variables $\Delta_x$, $Q_m$ and $\mathcal{C}_0$ are the optical detuning with the static optomechanical shift correction, $\Delta_x = \omega_l - \omega_0 - Gx_0$, the mechanical oscillator quality factor, $Q_m = \Omega_m/\Gamma_m$, and the single-photon cooperativity $\mathcal{C}_0 = 4g_0^2/\Gamma_m\kappa$, respectively. The derivatives now will be all with respect to $\widetilde{t}$ unless explicitly said the opposite. The motivations to construct a semi-analytical model in this article are (i) to prove that each term $F_n$ from the power expansion of the optical force is mainly responsible for the $p:q$ Arnold tongue width $\Delta\Omega(p,q)$, (ii) to obtain a semi-analytical formula for these $\Delta\Omega(p,q)$, (iii) to show that the influence of the optical detuning greatly change the synchronization region, (iv) to prove that the symmetry breaking term $F_2$, which is neglected in many articles, is actually crucial for the dynamic and (v) to explain/predict those sidebands around the synchronization region. We then start uncoupling \Cref{eq:optomechanics_3} using adiabatic considerations: our optomechanical cavity has mechanical linewidth $\Gamma_m$ much smaller than the optical linewidth $\kappa$, as well the mechanical frequency $\Omega_m$ also much smaller than $\kappa$, a regime called unresolved sidebands. We can then assume that $a(\widetilde{t})$ is always in equilibrium with $\widetilde{x}(\widetilde{t} - \widetilde{\tau})$, where $\widetilde{\tau}$ is some adimensional time delay that we will deduce later in \Cref{eq:tau}, so we can write $a(\widetilde{t})$ as

\vspace{-0.4cm}
\begin{equation}\label{eq:a_delayed}
    a(\widetilde{t}) \approx \sqrt{\frac{\kappa_e}{\kappa}}\frac{s(\widetilde{t})}{\sqrt{\kappa}}\frac{2}{1 -i\left[\frac{2\Delta_x}{\kappa} - \widetilde{x}(\widetilde{t}-\widetilde{\tau})\right]},
\end{equation}

\noindent and then we can analyze the whole system just looking to one equation

\vspace{-0.3cm}
\begin{equation}
    \ddot{\widetilde{x}}(\widetilde{t}) +\frac{\dot{\widetilde{x}}(\widetilde{t})}{Q_m} + \widetilde{x}(\widetilde{t}) = -\widetilde{x}_0 - \frac{\mathcal{C}_0}{Q_m}\left(\frac{2\kappa_e}{\kappa}\right)\left(\frac{2s_0^2}{\kappa}\right)\left(\frac{1 + \varepsilon\sin{\Theta_d(\widetilde{t})}}{1 + \left[\frac{2\Delta_x}{\kappa} - \widetilde{x}(\widetilde{t}-\widetilde{\tau})\right]^2}\right),
\end{equation}
\vspace{-0.3cm}

\noindent where we define $f(\widetilde{t})$ and $f_0$ as

\vspace{-0.4cm}
\begin{equation}
    f(\widetilde{t}) = f_0\left[1 + \varepsilon\sin{\Theta_d(\widetilde{t})}\right] \quad \text{and} \quad f_0 = \frac{\mathcal{C}_0}{Q_m}\left(\frac{2\kappa_e}{\kappa}\right)\left(\frac{2s_0^2}{\kappa}\right),
\end{equation}

\noindent which allows us to rewrite \Cref{eq:a_delayed} as

\vspace{-0.4cm}
\begin{equation}\label{eq:x_delayed}
    \ddot{\widetilde{x}}(\widetilde{t}) +\frac{\dot{\widetilde{x}}(\widetilde{t})}{Q_m} + \widetilde{x}(\widetilde{t}) = -\widetilde{x}_0 - \frac{f(\widetilde{t})}{1 + \left[\frac{2\Delta_x}{\kappa} - \widetilde{x}(\widetilde{t}-\widetilde{\tau})\right]^2},
\end{equation}

However, \Cref{eq:x_delayed} is still very complicated because it is a non-autonomous delay differential equation, so we will expand the RHS in a power series of $\widetilde{x}(\widetilde{t}-\widetilde{\tau})$ as

\vspace{-0.3cm}
\begin{equation}\label{eq:RHS_expansion}
    \frac{1}{1 + \left[\frac{2\Delta_x}{\kappa} - \widetilde{x}(\widetilde{t}-\widetilde{\tau})\right]^2} = F_{0} + F_{1}\widetilde{x}(\widetilde{t}-\widetilde{\tau}) + F_{2}\widetilde{x}^{2}(\widetilde{t}-\widetilde{\tau}) + F_{3}\widetilde{x}^{3}(\widetilde{t}-\widetilde{\tau}) + ...
\end{equation}
\vspace{0.0cm}

\noindent in which the actual form of these first coefficients (which were already shown in Fig. 1(b) of the article) are

\vspace{-0.4cm}
\begin{equation}\label{eq:coefficients}
    F_{0} = \frac{1}{1 + \frac{4\Delta_x^2}{\kappa^2}} \quad , \quad F_1 = \frac{2\left(\frac{2\Delta_x}{\kappa}\right)}{\left(1 + \frac{4\Delta_x^2}{\kappa^2}\right)^2} \quad , \quad F_2 = \frac{\left(\frac{12\Delta_x^2}{\kappa^2} - 1\right)}{\left(1 + \frac{4\Delta_x^2}{\kappa^2}\right)^3} \quad \text{and} \quad F_3 = \frac{4\left(\frac{2\Delta_x}{\kappa}\right)\left(\frac{4\Delta_x^2}{\kappa^2} - 1\right)}{\left(1 + \frac{4\Delta_x^2}{\kappa^2}\right)^4},
\end{equation}
\vspace{0.0cm}

\noindent which shows that large normalized detuning ($\Delta_x/\kappa\gg 1$) leads to negligible values, as each $F_{n+1}$ term decreases faster than $F_{n}$ as a function of $\Delta_x/\kappa$, i.e.,

\vspace{-0.4cm}
\begin{equation}
    \frac{F_{n}}{F_{n+1}} \sim \left(\frac{\Delta_x}{\kappa}\right).
\end{equation}

The value of each $F_n$ in our experiment was found to be $F_0=0.6711$, $F_1=0.6306$, $F_2=0.1421$ and $F_3=-0.2897$. Substituting \Cref{eq:RHS_expansion} in \Cref{eq:x_delayed} reveals the nonlinear nature of the optical feedback into the mechanical oscillator,

\vspace{-0.4cm}
\begin{equation}\label{eq:x_delayed_approx}
    \ddot{\widetilde{x}}(\widetilde{t}) +\frac{\dot{\widetilde{x}}(\widetilde{t})}{Q_m} + \widetilde{x}(\widetilde{t}) = -\widetilde{x}_0 - f(\widetilde{t})\left[F_{0} + F_{1}\widetilde{x}(\widetilde{t}-\widetilde{\tau}) + F_{2}\widetilde{x}^2(\widetilde{t}-\widetilde{\tau}) + F_{3}\widetilde{x}^3(\widetilde{t}-\widetilde{\tau})\right].
\end{equation}

To remove the delay dependence we can expand $\widetilde{x}(\widetilde{t}-\widetilde{\tau})$ in powers of $\widetilde{\tau}$ as

\vspace{-0.4cm}
\begin{equation}\label{eq:delay_expansion}
    \widetilde{x}^{n}(\widetilde{t}-\widetilde{\tau}) = \widetilde{x}^{n}(\widetilde{t}) - n\widetilde{\tau} \widetilde{x}^{n-1}(\widetilde{t})\dot{\widetilde{x}}(\widetilde{t}) + O(\widetilde{\tau}^2),
\end{equation}

\noindent where we are neglecting $O(\widetilde{\tau}^2)$ since we have that $\widetilde{\tau}^2$ is of order $O(\Omega_m^2/\kappa^2)$, as we will verify soon. We can then group all these terms in an arrangement that highlights how far from the ideal harmonic oscillator this system is, as shown in \Cref{eq:x_delayed_final}

\vspace{-0.3cm}
\begin{equation}\label{eq:x_delayed_final}
    \ddot{\widetilde{x}} +\left[\frac{1}{Q_m} - \widetilde{\tau}f(\widetilde{t})\left(F_{1} + 2F_{2}\widetilde{x} + 3F_{3}\widetilde{x}^2\right)\right]\dot{\widetilde{x}} + \left[1 + f(\widetilde{t})\left(F_{1} + F_{2}\widetilde{x} +F_{3}\widetilde{x}^2\right)\right]\widetilde{x} = -f_1(\widetilde{t})F_{0} - \left(f_0F_0+\widetilde{x}_0\right),
\end{equation}

\noindent where $f_1(\widetilde{t})$ is the AC component of $f(\widetilde{t})$, i.e., $f(\widetilde{t}) = f_0 + f_1(\widetilde{t})$ with $f_1(\widetilde{t}) = \varepsilon f_0\sin{\Theta_d(\widetilde{t})}$. The last term of \Cref{eq:x_delayed_final} in parentheses must be zero because is the only DC component of the whole equation and, as one can verify, solving $f_0F_0+\widetilde{x}_0=0$ returns the same static correction for the mechanical displacement $x(t)$ from the linearized optomechanics. \Cref{eq:x_delayed_final} is, finally, a well looking shape equation, very similar with Equation 8 from reference \cite{PhysRevE.91.032910} where their $\zeta$ and $f_\text{e}$ are related with our $F_1$ and $F_0$, besides we are not using temperature dynamics here. The generalization of \Cref{eq:x_delayed_final} until $O(\widetilde{\tau}^2)$ is given by

\vspace{-0.3cm}
\begin{equation}
    \ddot{\widetilde{x}} +\left[\frac{1}{Q_m} - \widetilde{\tau} f(\widetilde{t})\sum_{n=1}^{n_{\text{max}}}nF_n\widetilde{x}^{n-1}\right]\dot{\widetilde{x}} + \left[1 + f(\widetilde{t})\sum_{n=1}^{n_{\text{max}}}F_n\widetilde{x}^{n-1}\right]\widetilde{x} = -f_1(\widetilde{t})F_{0},
\end{equation}
\vspace{-0.1cm}

\noindent but we are not going to analyze this system, we will stick with the case $n_{\text{max}}=3$. We can do a final approximation which is to neglect terms of order $O(\widetilde{\tau}\varepsilon)$ as far $\varepsilon$ is kept small, knowing a priori that $\widetilde{\tau}$ is already small. We then have \Cref{eq:x_final}

\vspace{-0.4cm}
\begin{equation}\label{eq:x_final}
    \ddot{\widetilde{x}} +\left[\frac{1}{Q_m} - \widetilde{\tau} f_0\left(F_{1} + 2F_{2}\widetilde{x} + 3F_{3}\widetilde{x}^2\right)\right]\dot{\widetilde{x}} + \left[1 + \left(f_0 + f_1\right)\left(F_{1} + F_{2}\widetilde{x} +F_{3}\widetilde{x}^2\right)\right]\widetilde{x} = -f_1F_{0}
\end{equation}

\noindent and finally it is an ODE that all the terms proportional to $\widetilde{x}$ have a parametric excitation $f_1$, but terms proportional to $\dot{\widetilde{x}}$ do not. The only formula that is missing is $\widetilde{\tau} = \widetilde{\tau}(\Delta_x)$ for us to start studying \Cref{eq:x_final}. To find this missing expression note (i) that the term $f_0 F_1$ acts like a constant shift in the frequency, so we can associate with it the optical spring effect and (ii) that the term $-\widetilde{\tau} f_0 F_1$ acts like a constant change in the mechanical linewidth, so we can associate it with the optical cooling/heating. Doing that interconnection with the linearized optomechanical equations \cite{Aspelmeyer2014} we can identify an analytic expression for $\widetilde{\tau}$ because from our model we have that $\delta\Gamma_{m}^{\text{linear}} = -\widetilde{\tau} f_0 F_{1}$ and, from the linearized optomechanical equations,

\vspace{-0.3cm}
\begin{equation}\label{eq:gamma_lin}
    \delta\Gamma_{m}^{\text{linear}} = \frac{\mathcal{C}_0}{Q_m}\left(\frac{2\kappa_e}{\kappa}\right)\left(\frac{2s_0^2}{\kappa}\right)\left(\frac{1}{1 + \left(\frac{2\Delta_x}{\kappa}\right)^2}\right)\left(\frac{1}{1 + \left(\frac{2\Delta_x}{\kappa} + \frac{2\Omega_m}{\kappa}\right)^2} - \frac{1}{1+\left(\frac{2\Delta_x}{\kappa} - \frac{2\Omega_m}{\kappa}\right)^2}\right),
\end{equation}

\vspace{-0.3cm}

\begin{figure}[ht]
    \centering
    \includegraphics[width=0.95\linewidth]{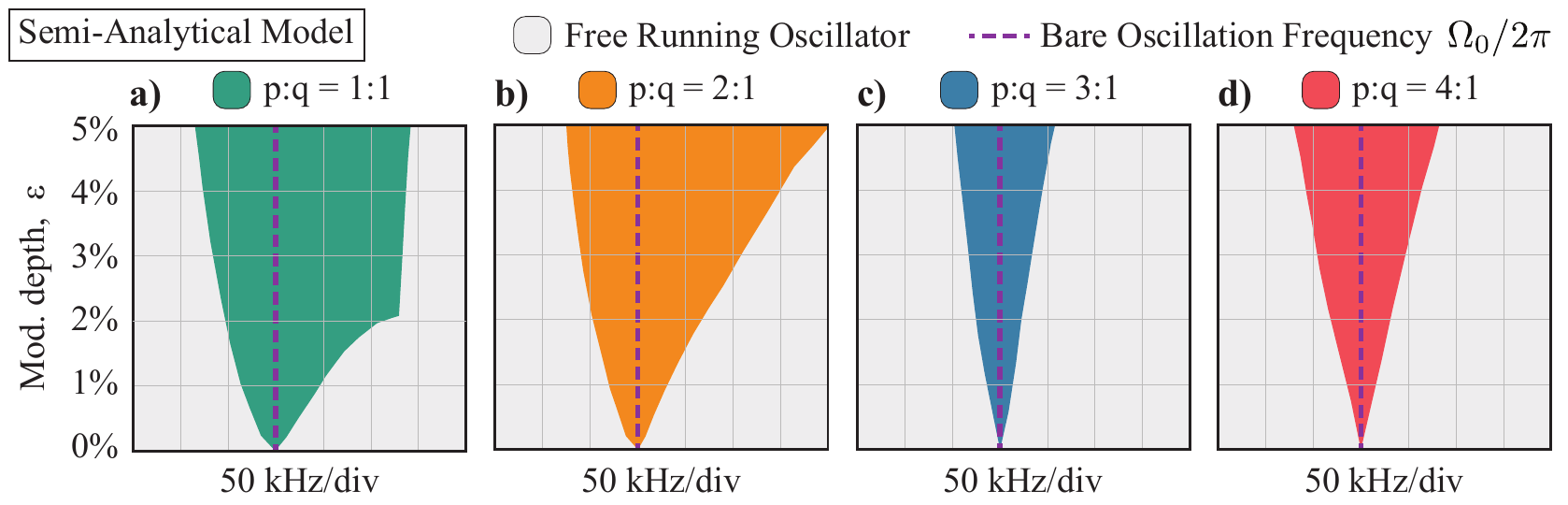}
    \caption{Simulated Arnold tongues using injection frequency $\Omega_d = p\Omega_{0}/q$ for the cases $p = \{1,2,3,4\}$ and $q=1$, in order, from \textbf{a)} to \textbf{d)}. To simulate these maps we used \Cref{eq:x_final}.}
    \label{fig:9}
\end{figure}

\begin{figure}[h!]
    \centering
    \includegraphics[width=\linewidth]{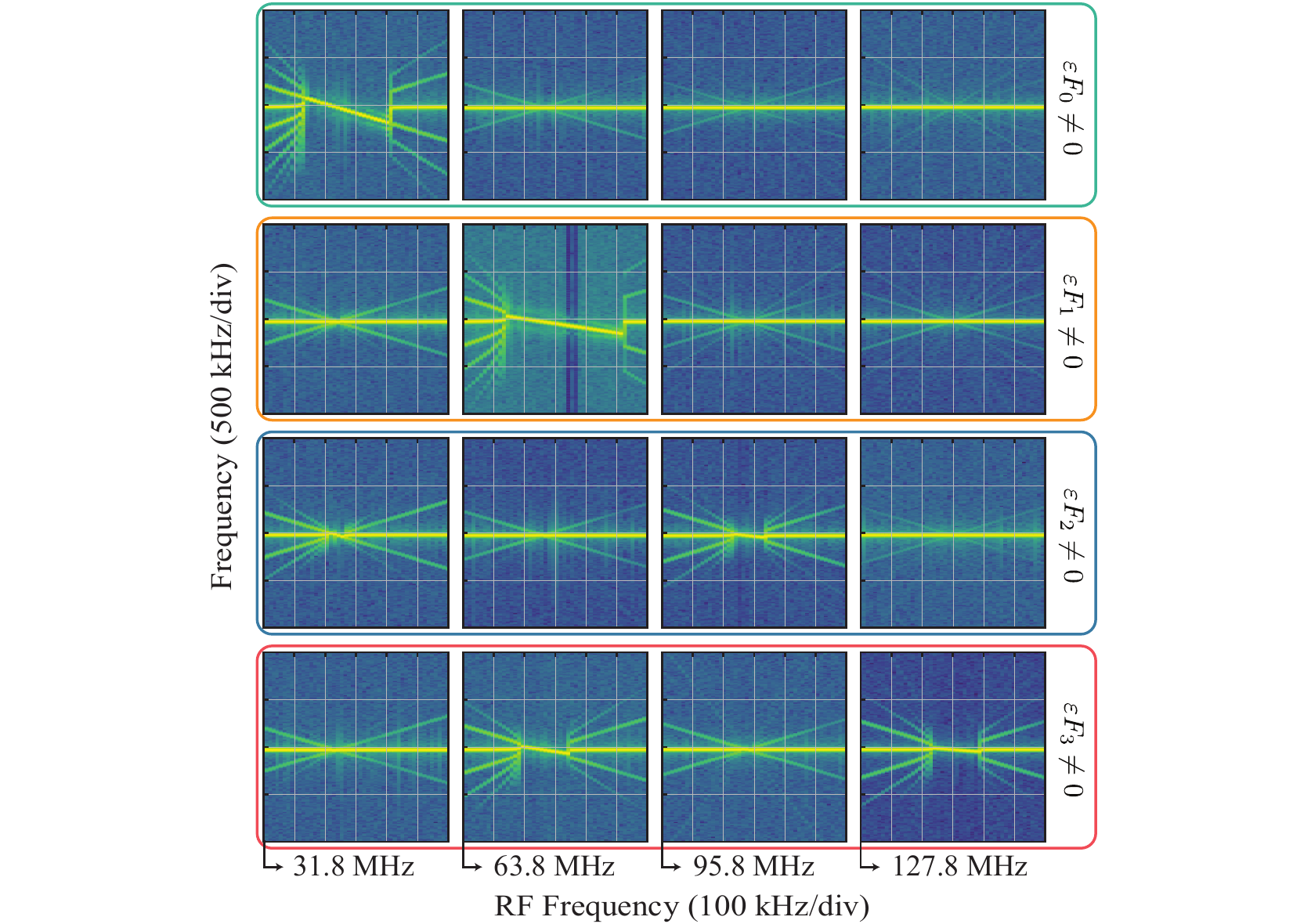}
    \caption{These simulations were done considering only one $\varepsilon F_n$ term in \Cref{eq:x_final}, e.g., the first row of this image-matrix-like consist of considering $\varepsilon F_1 = \varepsilon F_2 = \varepsilon F_3 = 0$ while $\varepsilon F_0 \neq 0 $ and doing the injection locking for all $p=1-4$. The same logic is valid for the others rows, as indicated in the right side of the figure. In these simulations we used $\varepsilon = 5\%$.}
    \label{fig:10}
\end{figure}

The Arnold tongues simulations using this semi-analytical model from \Cref{eq:x_final} are shown in \ref{fig:9}, and the comparison between this and \ref{fig:6} is striking. Before finishing this section we want to prove that the connection between $F_n$'s and the entire AT width $\Delta\Omega(n+1,1)$ is really strong and, for that, we simulated again \Cref{eq:x_final} but here we want to show not the AT map, but some particular spectrograms while we considered only one $F_n$ each time we simulate the system, as shown in \ref{fig:10}.

\newpage

And as we can see, simulating \Cref{eq:x_final} proves that the major dependence of each AT is indeed the parametric term $\varepsilon F_n$, as each one of these terms alone almost reproduces the whole dynamic of the system in a specific region. This hierarchical dependence is explicitly calculated, as shown in \Cref{semianalyticAT}.

\newpage

\section{}

\textbf{The averaging method of Krylov-Bogoliubov-Mitropolsky (KBM).} While the numerical simulation does predict many features and give us many insights about the observed data, it does not provide a direct prediction of the synchronization behavior. To pursue further analytical insight, we resort to the KBM method to derive amplitude and phase equations describing the coupling optomechanical oscillator enslaved by the driving modulated RF signal. We start here introducing new adimensional time $T$ and displacement $y$ given by

\vspace{-0.3cm}
\begin{equation}\label{eq:normalization_yT}
    \frac{\widetilde{x}}{y} = L_{\widetilde{x}} = -\frac{F_2}{3F_3} + \sqrt{\left(\frac{F_2}{3F_3}\right)^2 + \left(\frac{1-f_0F_1Q_m\widetilde{\tau}}{3f_0F_3Q_m\widetilde{\tau}}\right)} \quad \quad \text{and} \quad \quad \frac{\widetilde{t}}{T} = L_{\widetilde{t}} = \frac{1}{\sqrt{1 + f_0 F_1}}.
\end{equation}

At first glance, it seems like an awkward choice of normalization, but as we discuss below, they have a clear physical interpretation. \Cref{eq:normalization_yT} is the positive root of the coefficient of $\dot{\widetilde{x}}(\widetilde{t})$ in \Cref{eq:x_final}, which makes the amplitude of $y$ near the value of the limit circle of a van-der Pol oscillator $\approx O(1)$, i.e., we are renormalizing $\widetilde{x}$ by the positive solution of

\vspace{-0.5cm}
\begin{equation}\label{eq:x_dot_coef}
    \frac{1}{Q_m} - \widetilde{\tau} f_0\left(F_{1} + 2F_{2}\widetilde{x} + 3F_{3}\widetilde{x}^2\right) = 0.
\end{equation}

The choice of the new time scale makes the oscillation frequency of the oscillator, already accounted by the optical spring effect, about $\approx O(1)$. After these normalizations we have that \Cref{eq:x_final} becomes

\vspace{-0.3cm}
\begin{multline}\label{eq:master_y}
    \frac{d^2y}{dT^2}-\mu(1-y)(1+\sigma y)\frac{dy}{dT} + \left[1+\varepsilon\alpha\sin{\left(\omega T\right)}\right]y +\\
    + \beta\left[1+\varepsilon\sin{\left(\omega T\right)}\right]y^2 + \gamma\left[1+\varepsilon\sin{\left(\omega T\right)}\right]y^3 =  F\varepsilon\sin{\left(\omega T\right)},
\end{multline}

\noindent with new adimensional parameters defined as

\vspace{-0.4cm}
\begin{equation}
    \omega = \frac{\Omega_{d}}{\Omega_m}L_{\widetilde{t}}, \quad \quad \quad \mu = \left(\widetilde{\tau}f_0F_1Q_m - 1\right)L_{\widetilde{t}}, \quad \quad \quad \sigma = 1+\frac{2\widetilde{\tau}f_0F_2Q_m}{\widetilde{\tau}f_0F_1Q_m - 1}L_{\widetilde{x}},
\end{equation}

\vspace{-0.4cm}
\begin{equation*}
    \alpha = f_0F_1 L_{\widetilde{t}}^2, \quad \quad \quad \beta = f_0F_2 L_{\widetilde{x}}L_{\widetilde{t}}^2, \quad \quad \quad \gamma = f_0F_3L^2_{\widetilde{x}}L_{\widetilde{t}}^2, \quad \quad \text{and} \quad \quad F = -f_0F_0L_{\widetilde{t}}^2/L_{\widetilde{x}}.
\end{equation*}
\vspace{-0.3cm}

It is evident that every parametric term $\alpha$, $\beta$, $\gamma$ and $F$ is proportional to $f_0$, regardless of the convoluted terms $L_{\widetilde{t}}$ and $L_{\widetilde{x}}$, meaning that higher optical pump intensity enhance these terms. Also, each of these terms are proportional to one $F_n$, making clear distinction where each nonlinearity really is. Such model returns us the same used by Shreyas Y. Shah at \cite{PhysRevLett.114.113602} if $\mu = \gamma = F = 0$ and also neglecting the autonomous quadratic term $\beta y^2$ (which is the term that comes from a odd power potential, making the problem parity asymmetric), then

\vspace{-0.3cm}
\begin{equation}
    \frac{d^2y}{dT^2} + \left[1 + \alpha\varepsilon\sin{\left(\omega T\right)}\right]y + \beta\varepsilon\sin{\left(\omega T\right)}y^2 = 0,
\end{equation}

\noindent but to leave in the exact shape of the one used there we should change the independent variable $\omega T \rightarrow U + \frac{\pi}{2}$ and then

\vspace{-0.4cm}
\begin{equation}
    \frac{d^2y}{dU^2} + \left[\frac{1}{\omega^2} + \frac{\alpha}{\omega^2}\varepsilon\cos{\left(U\right)}\right]y + \frac{\beta}{\omega^2}\varepsilon\cos{\left(U\right)}y^2 = 0,
\end{equation}

\noindent from which we interpret the parameters as

\vspace{-0.4cm}
\begin{equation}
    \delta^{\text{Shah}} = \frac{1}{\omega^2} \quad , \quad D_1^{\text{Shah}} = \frac{\alpha}{\omega^2} \quad , \quad D_2^{\text{Shah}} = \frac{\beta}{\omega^2} \quad , \quad \gamma^{\text{Shah}} = \varepsilon.
\end{equation}

However, we will not use the multi-scale method to study synchronization neither to find bifurcations, our analysis will be based in the KBM method of averaging. The value of the parameters obtained from simulations are $\mu = 9.813 \times 10^{-4}$, $\sigma = 1.665 \times 10^{0}$, $\alpha = 2.383 \times 10^{-2}$, $\beta = 4.396 \times 10^{-3}$, $\gamma = -7.340 \times 10^{-3}$ and $F = -3.098 \times 10^{-2}$.

We begin our KBM analyzes from \Cref{eq:master_y}, which is a nonlinear oscillators of the form

\vspace{-0.3cm}
\begin{equation}\label{eq:y_oscillator_form}
    \frac{d^2y}{dT^2} + y = K\left(T, y,\frac{dy}{dT}\right),
\end{equation}

\noindent where $K$ is small compared to $y$. If $K(T,y,\frac{dy}{dT})=0$, we would have the ideal harmonic oscillator with solution $y = A\sin{\left(T+\Phi\right)}$ for any choice of constants $A$ and $\Phi$. If we now try solving \Cref{eq:y_oscillator_form} with slowly varying amplitude and phase $(A(T),\Phi(T))$ as ansatz, i.e.,

\vspace{-0.3cm}
\begin{equation}
    y = A(T)\sin{\left[T+\Phi(T)\right]}
    \quad \quad \text{and} \quad \quad
    \frac{dy}{dT} = A(T)\cos{\left[T+\Phi(T)\right]},
\end{equation}

\noindent we can show \cite{jackson1989perspectives} that this system has general solution given by \Cref{eq:pre_KBM}

\vspace{-0.2cm}
\begin{equation}\label{eq:pre_KBM}
    \begin{cases}
        \frac{dA}{dT} = \cos{\left(\phi\right)}K\left( T, A\sin{\phi}, A\cos{\phi}\right)
        \\
        \\
        \phi(T) = T+\Phi(T)
        \\
        \\
        \frac{d\Phi}{dT} = - \frac{\sin{\left(\phi\right)}}{A}K\left(T, A\sin{\phi}, A\cos{\phi}\right)
    \end{cases}
\end{equation}
\vspace{0.0cm}

The KBM method take its place here, where we average these equations over one period, however, the integral in $T$ is replaced over a integral in $\phi$ considering that $d\phi \approx dT$, which is correct to zero order in $\Phi(T)$, so

\vspace{-0.2cm}
\begin{equation}\label{eq:KBM_A}
    \left<\frac{dA}{dT}\right>_{T} \approx \left<\frac{dA}{dT}\right>_{\phi} = \frac{1}{2\pi}\int_{0}^{2\pi}\cos{\left(\phi\right)}K\left(\phi-\Phi, A\sin{\phi}, A\cos{\phi}\right)d\phi
\end{equation}

\begin{equation}\label{eq:KBM_Phi}
    \left<\frac{d\Phi}{dT}\right>_{T} \approx \left<\frac{d\Phi}{dT}\right>_{\phi} = -\frac{1}{2\pi A}\int_{0}^{2\pi}\sin{\left(\phi\right)}K\left(\phi-\Phi, A\sin{\phi}, A\cos{\phi}\right)d\phi
\end{equation}

\noindent and if our system were autonomous the integrals from \Cref{eq:KBM_A} and \Cref{eq:KBM_Phi} would be relatively easy to proceed, however, we have an external drive and this makes our system non-autonomous. To proceed the integral, we need to deal with $\Phi$, which we will just let constant during integration, arguing that $\Phi$ is a slow varying function of $T$. The general form of $K\left(T, y, \frac{dy}{dT}\right)$ for our system can be splitted in two contribution: one autonomous and another one non-autonomous, i.e.,

\vspace{-0.4cm}
\begin{equation}
    K\left(T, y, \frac{dy}{dT}\right) = K_{\text{auto}}\left(y, \frac{dy}{dT}\right) + K_{\text{non-auto}}\left(T, y, \frac{dy}{dT}\right)
\end{equation}

\noindent in which each part is written as

\vspace{-0.4cm}
\begin{equation}
    K_{\text{auto}}\left(y, \frac{dy}{dT}\right) = \mu\left(1-y\right)(1+\sigma y)\frac{dy}{dT} - \beta y^2 - \gamma y^3
\end{equation}

\vspace{-0.3cm}
\begin{equation}
    \text{and} \quad \quad K_{\text{non-auto}}\left(T, y, \frac{dy}{dT}\right) =\varepsilon\sin(\omega T)\left(F - \alpha y - \beta y^2 - \gamma y^3\right).
\end{equation}

Substituting $K\left(T, y, \frac{dy}{dT}\right)$ into \Cref{eq:KBM_A} and \Cref{eq:KBM_Phi} and then performing the integration over one period of $\phi$ we obtain our amplitude and phase differential equations given by

\vspace{0.0cm}
\begin{multline}\label{eq:KBM_A_final}
    \left<\frac{dA}{dT}\right>_{\phi} = \frac{\mu A}{2}\left(1-\frac{\sigma A^2}{4}\right) +\\ 
    + \frac{\varepsilon\sin (\pi\omega)}{\pi}\left(\frac{\omega \left[F \left(\omega ^2-9\right) + 2\beta A^2\right]\sin{\left[(\pi -\Phi ) \omega\right]}}{\left(\omega
   ^2-9\right) \left(\omega ^2-1\right)} - \frac{A \left[\alpha  \left(\omega ^2-16\right)-6 \gamma A^2 \right] \cos{\left[(\pi -\Phi ) \omega \right]}}{\left(\omega
   ^2-16\right) \left(\omega ^2-4\right)}\right),
\end{multline}

\begin{multline}\label{eq:KBM_Phi_final}
    \left<\frac{d\Phi}{dT}\right>_{\phi} = \frac{3\gamma A^2}{8} +\\
    -\frac{\varepsilon\sin(\pi\omega)}{\pi}\left(\frac{\left[F \left(\omega ^2-9\right) + 6\beta A^2\right]\cos{\left[(\pi -\Phi ) \omega\right]}}{\left(\omega
   ^2-9\right) \left(\omega ^2-1\right)A} - \frac{2 \left[\alpha  \left(\omega ^2-16\right)-12 A^2 \gamma \right]\sin{\left[(\pi -\Phi ) \omega \right]}}{\omega
   \left(\omega ^2-16\right) \left(\omega ^2-4\right)}\right).
\end{multline}
\vspace{0.0cm}

This averaging technique is the essence of the KBM method to obtain amplitude and phase equations of nonlinear oscillators. Before we proceed, we can study \Cref{eq:KBM_A_final} and \Cref{eq:KBM_Phi_final} for the case $\varepsilon = 0$, which would give us exact solutions for both $A(T)$ and $\Phi(T)$ as

\vspace{-0.3cm}
\begin{equation}\label{subsec:KBM_infty}
    \begin{cases}
        A(T) = \frac{\pm 2}{\sqrt{\sigma + \left(\frac{4}{A_0} - \sigma\right) e^{-\mu T}}} \quad \Rightarrow \quad \lim_{T \rightarrow \infty}A(T) = A_{\infty} = \pm\frac{2}{\sqrt{\sigma}}
        \\
        \\
        \Phi(T) = \Phi_0 + \frac{3\gamma}{2\mu\sigma}\ln{\left[1 + \frac{\sigma A_0^2}{4}\left(e^{\mu T} - 1\right)\right]} \quad \Rightarrow \quad \lim_{T \rightarrow \infty}\Phi(T) = \Phi_{\infty} = \frac{3\gamma}{2\sigma}T
    \end{cases}
\end{equation}
\vspace{0.0cm}

\noindent for constants $A_0$ and $\Phi_0$. Here we conclude that even with zero modulation depth there exist a frequency shift contribution that comes from the Duffing term $\gamma$. The steady oscillation frequency $\Omega_0$ is given by

\vspace{-0.3cm}
\begin{equation}
    \Omega_0(\varepsilon=0) = \lim_{t \rightarrow \infty}\frac{d\phi}{dt} = \lim_{t \rightarrow \infty}\frac{d\phi}{dT}\frac{dT}{dt} = \frac{\Omega_m}{L_{\widetilde{t}}}\frac{d}{dT}\left(T + \Phi_{\infty}\right) = \Omega_m\sqrt{1 + f_0 F_1}\left(1 + \frac{3\gamma}{2\sigma}\right) 
\end{equation}
\vspace{-0.1cm}

\noindent which can be used to estimate the Duffing term from the measured oscillation frequency. Remembering the definitions of $\omega$ and $\rho$ we can obtain a really nice relation between them as

\vspace{-0.3cm}
\begin{equation}
 \omega = \frac{\Omega_d}{\Omega_m\sqrt{1 + f_0 F_1}} \quad \quad \& \quad \quad \rho = \frac{p}{q} = \frac{\Omega_d}{\Omega_0} \quad \quad \Rightarrow \quad \quad \frac{\omega}{\rho} = 1 + \frac{3\gamma}{2\sigma} + O(\varepsilon),
\end{equation}

\noindent and its clear now that if $\rho$ is an integer, it does not mean that $\omega$ is an integer. In our case (and most of the cases) the Duffing correction in the frequency is really small, as a matter of fact we have

\vspace{-0.3cm}
\begin{equation}
    \frac{3\gamma}{2\sigma} \approx -6.6125 \times 10^{-3} = O(10^{-3}) \ll O(1)
\end{equation}

\noindent so we will considerer $\omega \approx \rho$ for now on, because that is an excellent approach to obtain simple, but good, analytical results. The phase equation can be expanded in the vicinity of some integer $\rho = \{1,2,3,4\}$ to give us insight into higher harmonic synchronization. We have then the following cases

\vspace{-0.3cm}
\begin{equation}\label{subsec:simulated_tongues_KBM}
    \begin{cases}
        \lim_{\rho \rightarrow 1}\left<\frac{d\Phi}{dT}\right>_{\phi} \approx \frac{3\gamma A^2}{8} - \varepsilon\left(\frac{F}{2A} - \frac{3\beta A}{8}\right)\cos{\left(\Phi\right)},
        \\
        \\
        \lim_{\rho \rightarrow 2}\left<\frac{d\Phi}{dT}\right>_{\phi} \approx \frac{3\gamma A^2}{8} +\varepsilon\left(\frac{\alpha}{4}+\frac{\gamma A^2}{4}\right)\sin{\left(2\Phi\right)}.
        \\
        \\
        \lim_{\rho \rightarrow 3}\left<\frac{d\Phi}{dT}\right>_{\phi} \approx \frac{3\gamma A^2}{8} -\varepsilon\frac{\beta A}{8}\cos{\left(3\Phi\right)}.
        \\
        \\
        \lim_{\rho \rightarrow 4}\left<\frac{d\Phi}{dT}\right>_{\phi} \approx \frac{3\gamma A^2}{8} -\varepsilon\frac{\gamma A^2}{16}\sin{\left(4\Phi\right)}.
    \end{cases}
\end{equation}

When the system is locked to the driving signal, we know that the amplitude $A(T)$ of the oscillator is almost constant (this is the Kuramoto approximation) and that the phase $\phi(T) = T + \Phi(T)$ is a linear function of time $T$ because, otherwise, the oscillation frequency $\Omega_0$ would not be static, i.e., it would fluctuate around some mean frequency. In other words, we are imposing that the derivative of $\phi(T)$ to be constant during the locking, so we can say that $\Phi(T) = \delta_{\rho} T$ for some frequency mismatch $\delta_{\rho}$ of our bare oscillator, i.e.,

\vspace{-0.5cm}
\begin{equation}
    \phi(T) = (1 + \delta_\rho)T \quad \Rightarrow \quad \frac{d\phi}{dT} = 1 + \delta_\rho,
\end{equation}

\noindent and then we can solve for $\varepsilon = \varepsilon(\delta_\rho)$ for various $\delta_\rho$. The cases $\rho = 1$ and $\rho = 2$ are

\vspace{-0.0cm}
\begin{equation}\label{eq:AT_boundary}
    \begin{cases}
        \delta_{1} = \frac{3\gamma A^2}{8} - \varepsilon\left(\frac{F}{2A} - \frac{3\beta A}{8}\right)\cos{\left(\delta_{1}T\right)},
        \\
        \\
        \delta_{2} =  \frac{3\gamma A^2}{8} +\varepsilon\left(\frac{\alpha}{4}+\frac{\gamma A^2}{4}\right)\sin{\left(2\delta_{2}T\right)},
    \end{cases}
    \quad \Rightarrow \quad
    \begin{cases}
        \varepsilon(\delta_1) > \left|\left(\delta_{1} - \frac{3\gamma A^2}{8}\right)/\left(\frac{F}{2A} - \frac{3\beta A}{8}\right)\right|,
        \\
        \\
        \varepsilon(\delta_2) > \left|\left(\delta_{2} - \frac{3\gamma A^2}{8}\right)/\left(\frac{\alpha}{4}+\frac{\gamma A^2}{4}\right)\right|,
    \end{cases}
\end{equation}

\begin{figure}[ht]
    \centering
    \includegraphics[width=\linewidth]{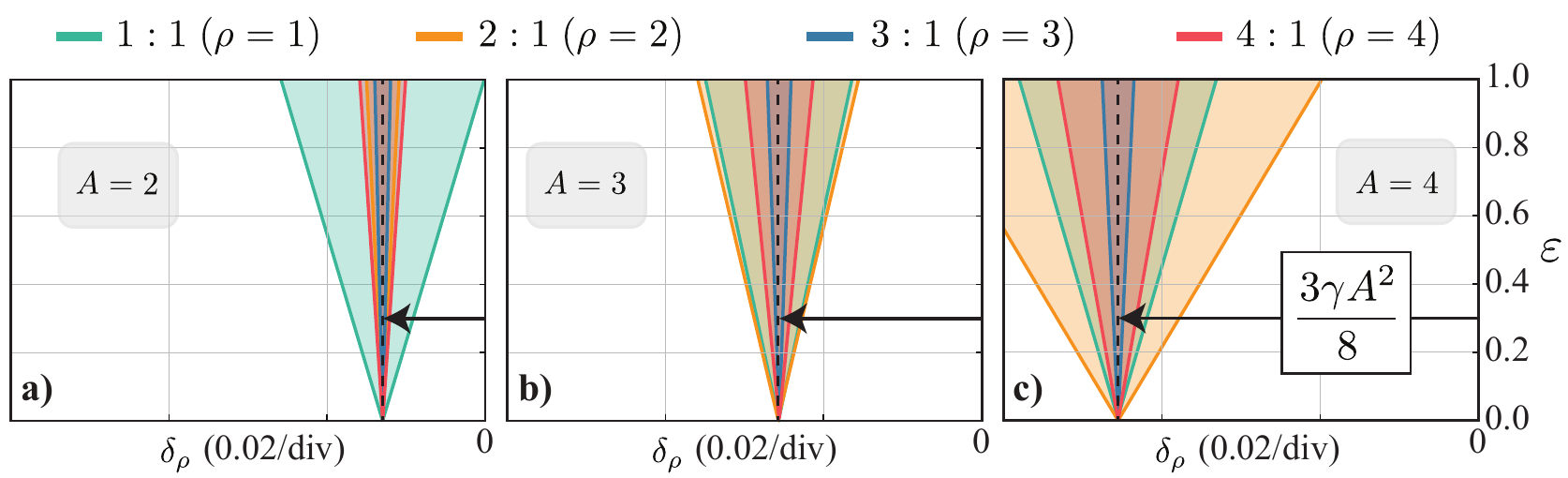}
    \caption{Simulated Arnold tongues using \Cref{subsec:simulated_tongues_KBM} for three different oscillations amplitude. \textbf{a)} $A=2$; \textbf{b)} $A=3$; \textbf{c)} $A=4$. The softening Duffing effect is enhanced as we increase the amplitude $A$, as we can verify as the whole region is getting away from $\delta_\rho = 0$ to the left, i.e., negative values of $\delta_\rho$.}
    \label{fig:11}
\end{figure}

\newpage

\noindent which defines a region in a $\varepsilon-\delta_\rho$ space which is, as we would guess, the Arnold tongues. The AT maps using this approach are shown in \ref{fig:11} for three different oscillation amplitudes $A = \{2,3,4\}$, which has the same effects of ones obtained from the experiment even after lots of approximations.

The reduction of the oscillation frequency $\Omega_{0}$ as we increase the amplitude $A$ is called softening Duffing, which is expected because we have $\gamma < 0$ (a consequence of the chosen optical detuning $0 < \Delta_x < \kappa/2$). If we had chosen other detuning $\Delta_x$, for example, $\Delta_x > \kappa/2$ or $\Delta_x < 0$, we would have $\gamma > 0$ and we would see the hardening Duffing effect, which is the shift of $\Omega_{0}$ to higher frequencies. Not only this but for high enough amplitudes we can obtain larger $2:1$ AT than $1:1$, which is the case for $A=3$ and $A=4$, showing that our model still has features about high-harmonic synchronization, the same features of the experiment. To finalize this section, we conclude that each of the terms $F, \alpha, \beta$ and $\gamma$ are directly proportional to the tongue width $\Delta\Omega(p,q)$ with $p = \{1,2,3,4\}$ and $q=1$, respectively, as we can see at the denumerator of \Cref{eq:AT_boundary}, i.e.,

\begin{equation}\label{semianalyticAT}
    \begin{cases}
        \Delta\Omega(1,1) = 2\times\left(\frac{F}{2A}-\frac{3\beta A}{8}\right) \approx \frac{F}{A} \propto F \propto F_0
        \\
        \\
        \Delta\Omega(2,1) = 2\times\left(\frac{\alpha}{4}+\frac{\gamma A^2}{4}\right) \approx \frac{\alpha}{2} \propto \alpha \propto F_1
        \\
        \\
        \Delta\Omega(3,1) = 2\times\frac{\beta A}{8} \propto \beta \propto F_2
        \\
        \\
        \Delta\Omega(4,1) = 2\times\frac{\gamma A^2}{16} \propto \gamma \propto F_3
    \end{cases}
\end{equation}
\vspace{0.2cm}

\noindent and these are the semi analytical expressions for the tongue width of each harmonic, which we could engineer it to achieve wider Arnold tongues for different harmonics choosing different $F$, $\alpha$, $\beta$ and $\gamma$ as we design the geometry and the materials of our optomechanical cavity.

\vspace{-0.3cm}
\section{}

\textbf{Sidebands around the carrier at the synchronized region.} To explain the sidebands around the synchronization region we will first linearize \Cref{eq:KBM_A_final} and \Cref{eq:KBM_Phi_final} expanding $A(T)$ and $\Phi(T)$ as

\vspace{-0.5cm}
\begin{equation}\label{Sidebands:expansion}
    A(T) = \overline{A} + \delta A(T) \quad \quad \text{and} \quad \quad \Phi(T) = \overline{\Phi} + \delta \Phi(T),
\end{equation}
\vspace{-0.5cm}

\noindent and then diagonalize the linear part of the system

\vspace{-0.4cm}
\begin{equation}\label{Sidebands:linear}
    \frac{d}{dT}
    \begin{pmatrix}
        \delta A \\
        \delta \Phi
    \end{pmatrix}
    =
    \begin{pmatrix}
        H_{AA} & H_{A\Phi}\\
        H_{\Phi A} & H_{\Phi \Phi}
    \end{pmatrix}
    \begin{pmatrix}
        \delta A \\
        \delta \Phi
    \end{pmatrix},
\end{equation}

\noindent in which the actual form of $H_{AA}$, $H_{A\Phi}$, $H_{\Phi A}$ and $H_{\Phi \Phi}$ are too big to be shown and not important for our present analysis. The eigenvalues of this system give us the first order correction in frequency and damping of our oscillator. The evolution of this system is

\vspace{-0.2cm}
\begin{equation}
    \begin{pmatrix}
    \delta A(T)\\
    \delta \Phi(T)
    \end{pmatrix}
    =
    \begin{pmatrix}
    \delta A_+\\
    \delta \Phi_+
    \end{pmatrix}
    e^{\left(\lambda_{\text{Re}}^{+} + i\lambda_{\text{Im}}^{+}\right)T}
    +
    \begin{pmatrix}
    \delta A_-\\
    \delta \Phi_-
    \end{pmatrix}
    e^{\left(\lambda_{\text{Re}}^{-} + i\lambda_{\text{Im}}^{-}\right)T},
\end{equation}

\noindent where $\lambda^{\pm} = \lambda_{\text{Re}}^{\pm} + i\lambda_{\text{Im}}^{\pm}$ are the eigenvalues. Using these in $y(T)$ we can now search for sidebands, i.e.,

\vspace{-0.4cm}
\begin{multline}
    y(T) = A(T)\sin{\left[T + \Phi(T)\right]} \approx \frac{1}{2i}\left(\overline{A} + \delta A\right)\left[\left(1+i\delta \Phi\right)e^{iT}e^{i\overline{\Phi}} - \left(1-i\delta \Phi\right)e^{-iT}e^{-i\overline{\Phi}}\right] \approx \\
    \approx \overline{A}\sin{\left(T+\overline{\Phi}\right)} + \delta{A}(T)\sin{\left(T+\overline{\Phi}\right)} + \overline{A}\delta\Phi(T)\cos{\left(T+\overline{\Phi}\right)}.
\end{multline}
\vspace{-0.3cm}

It is evident now that our oscillator has more than one single frequency because of the product $\delta A(T)$ with the sine function and also because of the product $\delta\Phi(T)$ with the cosine. The frequencies $\Omega_{\text{SB}}^{\pm}$ and the linewidths $\Gamma_{\text{SB}}^{\pm}$ of these new sidebands are given by

\begin{figure}[ht]
    \centering
    \includegraphics[width=\linewidth]{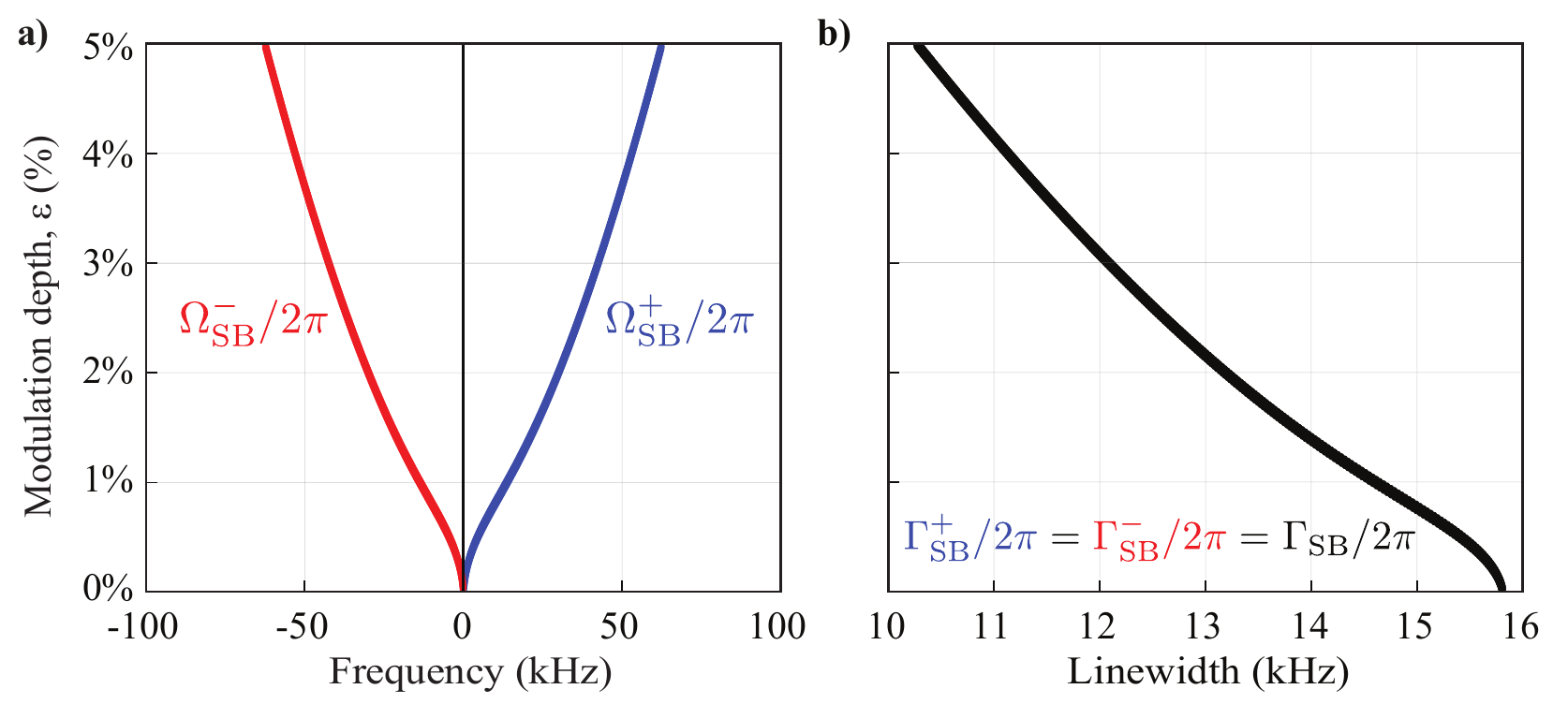}
    \caption{\textbf{a)} Detuning from $\Omega_0/2\pi$ showing both sidebands emerging as we increase the modulation depth; \textbf{b)} Linewidth of these sidebands.}
    \label{fig:12}
\end{figure}

\vspace{-0.4cm}
\begin{equation}
    \Omega_{\text{SB}}^{\pm} = \frac{\lambda_{\text{Im}}^{\pm}}{L_{\widetilde{t}}} \quad \quad \text{and} \quad \quad \Gamma_{\text{SB}}^{\pm} = \frac{\lambda_{\text{Re}}^{\pm}}{L_{\widetilde{t}}},
\end{equation}
\vspace{-0.1cm}

\noindent and the graph of these using our semi-analytical model for the case $\rho = 1$ is shown in \ref{fig:12}, which has an excellent agreement with the experimental data in both frequency and linewidth, showing us that these sidebands are indeed a coupling between phase and amplitude of the mechanical oscillator. 

\vspace{-0.3cm}
\section{}

\begin{figure}[h!]
    \centering
    \includegraphics[width=\linewidth]{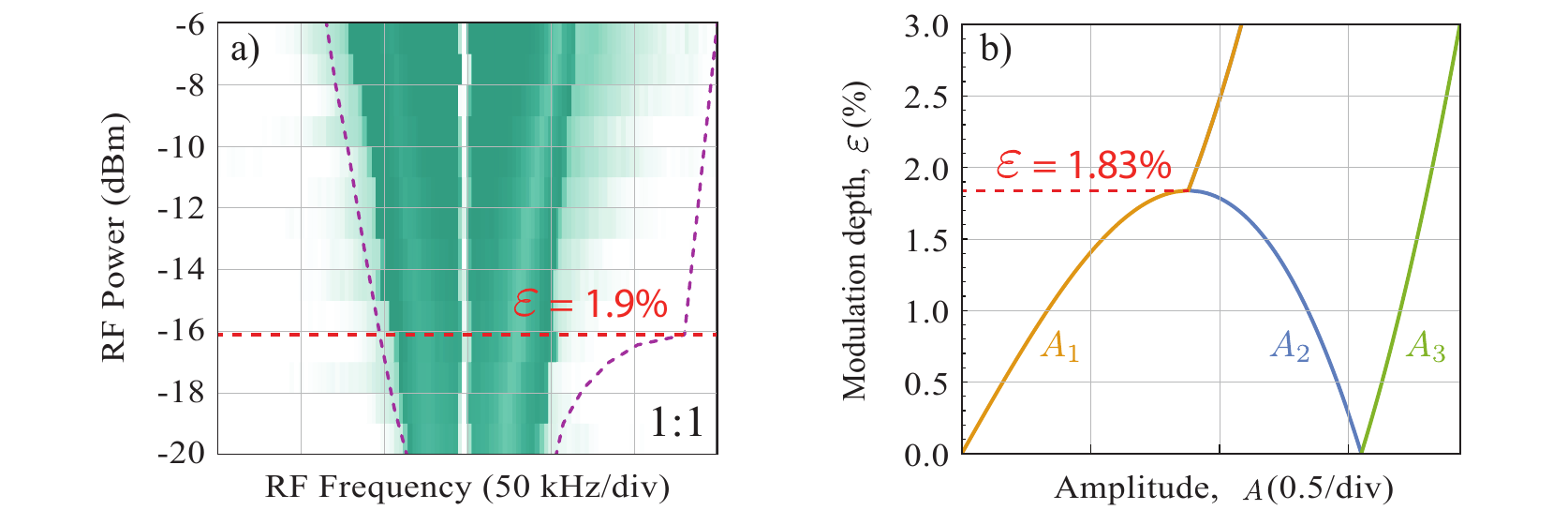}
    \caption{\textbf{a)} Experimental $1:1$ Arnold tongue with its simulated boundary in purple. The cusp occurs around -16 dBm, which corresponds to $\varepsilon = 1.9\%$; \textbf{b)} Amplitude evolution of each $A_n$ root from \Cref{eq:A_roots} as the modulation depth $\varepsilon$ increases.}
    \label{fig:14}
\end{figure}

\textbf{1:1 Arnold Tongue Cusp Deviation.} The purple cusp simulation in Fig.2 (d3) from the main text (also shown again in \ref{fig:14}(a)), which is not present in the experimental data, was found to be a bifurcation in the amplitude dynamics. To show this we must investigate \Cref{eq:KBM_A_final} in more detail:

\vspace{-0.2cm}
\begin{multline}
    \left<\frac{dA}{dT}\right>_{\phi} = \frac{\mu A}{2}\left(1-\frac{\sigma A^2}{4}\right) +\\ 
    + \frac{\varepsilon\sin (\pi\omega)}{\pi}\left(\frac{\omega \left[F \left(\omega ^2-9\right) + 2\beta A^2\right]\sin{\left[(\pi -\Phi ) \omega\right]}}{\left(\omega
   ^2-9\right) \left(\omega ^2-1\right)} - \frac{A \left[\alpha  \left(\omega ^2-16\right)-6 \gamma A^2 \right] \cos{\left[(\pi -\Phi ) \omega \right]}}{\left(\omega
   ^2-16\right) \left(\omega ^2-4\right)}\right).
\end{multline}

We are searching for bifurcations around the $1:1$ Arnold tongue, so it's natural to expand such equation around $\omega = 1$, identically as we've done with the phase $\Phi$ in  \Cref{subsec:simulated_tongues_KBM}, i.e.,

\begin{equation}
    \left<\frac{dA}{dT}\right>_{\phi} = \frac{\mu A}{2}\left(1-\frac{\sigma A^2}{4}\right) - \varepsilon\left(\frac{F}{2}-\frac{\beta A^2}{8}\right)\sin{\left[\Phi(T)\right]}.
\end{equation}

The fixed points of this equation are such that $\dot{A} = \dot{\Phi} = 0$. As the bifurcation occurs at the boundary of the Arnold tongue, we will assume a fixed phase $\Phi$ of $\pi/2$ and $3\pi/2$ (which, respectively, minimize and maximize $\sin{\Phi}$), such that the equation that we must solve for $A$ is

\begin{equation}\label{eq:A_roots}
    0 = \frac{\mu A}{2}\left(1-\frac{\sigma A^2}{4}\right) \pm \varepsilon\left(\frac{F}{2}-\frac{\beta A^2}{8}\right).
\end{equation}

We know that every $n$-degree polynomial has $n$ complex roots, so we can plot the absolute value of each of these roots as a function of $\varepsilon$ (we will use the same $\mu$, $\sigma$, $F$, and $\beta$ parameters from the simulation), and the result is shown in \ref{fig:14}(b)

As we see, two of the roots ($A_1$ and $A_2$) of \Cref{eq:A_roots} become degenerate (the orange and blue curves of \ref{fig:14}(b)) for a modulation depth around $\varepsilon = 1.83\%$. The cusp that we have at the \ref{fig:14}(a) happens around -16 dBm, which corresponds to a modulation depth of $1.9\%$. We must not forget that \Cref{eq:A_roots} is one of the final steps of our semi-analytical analysis, and many simplifications were  involved. Nevertheless, the purple curve in \ref{fig:14} was obtained using the full optomechanical model, and a difference of only $0.07\%$ in $\varepsilon$ is observed. Similar cusps were observed in other injection-locking experiments \cite{PhysRevE.50.3383}. A more detailed analysis and more experimental data should be gathered to understand why our experiment does not exhibit this feature.

\bibliographystyle{plain}
\section*{Supplementary references}
\bibliography{supp_refs.bib}